\documentclass[11pt]{article}
\usepackage[protrusion=true,expansion=true]{microtype}
\usepackage{amsmath,amsfonts, braket}
\usepackage{tikz}
\usetikzlibrary{trees,er,snakes,shapes,mindmap}
\usepackage{titlesec}
\titleformat{\section}
 {\normalfont\fontfamily{bch}\fontsize{12pt}{16pt}\bfseries\color{black}}
{\thesection}{1em}{}
\def \beq  {\begin{equation}}
\def \eeq  {\end{equation}}
\def \beqar {\begin{eqnarray}}
\def \eeqar {\end{eqnarray}}
\allowdisplaybreaks
\def\sqr#1#2{{\vcenter{\vbox{\hrule height.#2pt
\hbox{\vrule width.#2pt height#1pt \kern#1pt
\vrule width.#2pt}\hrule height.#2pt}}}}

\def\la {{\langle}}
\def\ra {{\rangle}}

\def\a {\alpha}
\def\b {\beta}
\def\g {{\gamma}}
\def\d {{\delta}}

\def\d {\delta}

\def\bz {{\bar{z}}}

\def\D {{\cal D}}

\def\half{\textstyle{1\over 2}}

\mathchardef\mhyphen="2D
\begin{document}
\fontfamily{bch}\fontsize{11pt}{15pt}\selectfont
\def \CMP {{Commun. Math. Phys.}}
\def \PRL {{Phys. Rev. Lett.}}
\def \PL {{Phys. Lett.}}
\def \NPBProc {{Nucl. Phys. B (Proc. Suppl.)}}
\def \NP {{Nucl. Phys.}}
\def \RMP {{Rev. Mod. Phys.}}
\def \JGP {{J. Geom. Phys.}}
\def \CQG {{Class. Quant. Grav.}}
\def \MPL {{Mod. Phys. Lett.}}
\def \IJMP {{ Int. J. Mod. Phys.}}
\def \JHEP {{JHEP}}
\def \PR {{Phys. Rev.}}
\def \JMP {{J. Math. Phys.}}
\def \GRG{{Gen. Rel. Grav.}}
\begin{titlepage}
\null\vspace{-62pt} \pagestyle{empty}
\begin{center}
\vspace{1.3truein} 
{\Large\bf
Entanglement entropy for integer quantum Hall effect}\\
~ \\
{\Large\bf in two and higher dimensions}
\vskip .5in
{\Large\bfseries ~}\\
 {\text\large{\bf Dimitra Karabali}}\\
\vskip .2in
{\sl  Physics and Astronomy Department,
Lehman College, CUNY\\
Bronx, NY 10468}\\
 \vskip .1in
\begin{tabular}{r l}
E-mail:&\!\!\!{\fontfamily{cmtt}\fontsize{11pt}{15pt}\selectfont dimitra.karabali@lehman.cuny.edu}\\
\end{tabular}
\vskip 1.5in
\centerline{\large\bf Abstract}
\end{center}
We analyze the entanglement entropy, in real space, for the higher dimensional integer quantum Hall effect on ${\mathbb {CP}}^k$ (any even dimension) for abelian and nonabelian magnetic background fields. In the case of $\nu=1$ we perform a semiclassical calculation which gives the entropy as proportional to the phase-space area. This exhibits a certain universality in the sense that the proportionality constant is the same for any dimension and for any background, abelian or nonabelian. We also point out some distinct features in the profiles of the eigenfunctions of the two-point correlator that underline the difference in the value of entropies between $\nu=1$ and higher Landau levels. 
\end{titlepage}
\pagestyle{plain} \setcounter{page}{2}
\section{Introduction}

Entanglement has been used to explore properties of quantum states in a variety of condensed matter systems. Typically a system is divided into two subsystems and the entanglement is calculated in terms of the von Neumann entropy of the reduced density matrix of one of the subsystems. For gapped two-dimensional systems, the leading order contribution to the entanglement entropy is proportional to the perimeter of the boundary separating the two subsystems, in particular $S= c L + \gamma + {\cal{O}}(1/L)$, where $L$ is the length of the boundary, $c$ is a non-universal coefficient and $\gamma$ is a universal quantity called topological entanglement entropy \cite{kitaev}. 

Of particular interest among two-dimensional gapped systems are the quantum Hall systems whose entanglement entropy has been widely studied under different partitions. For a real-space partition $\gamma=0$ for fully filled integer Quantum Hall states and nonzero for fractional quantum Hall states \cite{Sierra}-\cite{krempa}. The entanglement entropy in the case of integer quantum Hall states is amenable to analytical calculations due to the fact that the many-body ground state is in terms of free fermions. The area-law entropy behavior for the two-dimensional integer QHE was studied in different geometries analytically for $\nu=1$ and numerically up to $\nu=5$ in \cite{Sierra} and the coefficient $c$ was identified in these cases. 

In this paper we extend the calculation of the entanglement entropy in the case of higher dimensional integer quantum Hall effect (any even dimension), in particular quantum Hall  effect on ${\mathbb {CP}}^k$ \cite{KN}-\cite{Kar1}. For $k=1$ this reduces to the well known case of QHE on $S^2$ where the magnetic field is created by a monopole at the center \cite{Haldane}. The formulation of QHE on  ${\mathbb {CP}}^k$ for $k >1$ displays two interesting features: higher dimensionality and the possibility of introducing both abelian and nonabelian magnetic fields. In the latter case one deals with a many-body system of free fermions with internal degrees of freedom which is amenable to analytical calculations. 

The paper is organized as follows. In section 2 we give a brief description of the integer quantum Hall effect on ${\mathbb {CP}}^k$, the corresponding spectrum and the Landau level wavefunctions. In section 3 we focus on the lowest Landau level and derive analytical expressions for the entropy for a) arbitrary $k$ and abelian $U(1)$ magnetic field and b) $k=2$ for $U(1) \times SU(2)$ nonabelian magnetic field with fermions in the triplet representation. The entropy is expressed in terms of the eigenvalues of the two-point correlation function.  We perform a semiclassical calculation for the entropy and find that the area law as expressed in terms of a ``phase-space" area has the same coefficient $c$ for $\nu=1$ for any dimension and any abelian or nonabelian background. In section 4 we derive analytical expressions for the eigenvalues of the two-point correlator for the first Landau level and the $\nu=2$ quantum Hall system on $S^2$ and comment on how the different profiles account for the different values of $c$ in each case. We end with a short discussion.

\section { Quantum Hall effect on ${\mathbb {CP}}^k$}

In this section we will give a brief description of the Landau level states and wavefunctions for the quantum Hall effect on ${\mathbb {CP}}^k$, following a group theoretic analysis developed in \cite{KN}-\cite{Kar1}. ${\mathbb {CP}}^k$ is a $2k$-dimensional manifold which can be thought of as a coset space,
\beq
{\mathbb{CP}^k} = {SU(k+1) \over U(k)}  
\label{coset}
\eeq
The Landau wavefunctions can be obtained as functions of $SU(k+1)$ with specific transformation properties under the $U(k)$ subgroup. A basis for such functions is given by the so-called Wigner $\cal{D}$-functions, which are the matrices corresponding to the group elements in the unitary irreducible representations, namely
\beq
\D^{J}_{L,R}(g) = \la J , l_A \vert\, g\,\vert J, r_A \ra \label {wigner}
\eeq
where $J$ denotes the representation
and $l_A, ~r_A$ stand for two sets of quantum numbers specifying the 
states within the representation. On an element $g\in SU(k+1)$, 
we can define left and right $SU(k+1)$ actions by
\beq
{\hat{L}}_A ~g = T_A ~g, \hskip 1in {\hat{R}}_A~ g = g~T_A
\eeq
where $T_A$ are the $SU(k+1)$ generators in the representation to which $g$ belongs. The left transformations correspond to magnetic translations. There are $2k$ right generators of $SU(k+1)$ which are not in
$U(k)$; these can be separated into $T_{+i}$, $i=1,2 \cdots ,k$, which are
of the raising type and $T_{-i}$ which are of the lowering
type. These generate translations while $U(k)$ generates rotations at a point. The covariant derivatives on ${\mathbb {CP}}^k$ are given by 
\beq
\D_{\pm i}  = i\,{{\hat R}_{\pm i} \over r}
\label{covariant}
\eeq
where $r$ can be thought of as the radius of ${\mathbb {CP}}^k$. 
This is consistent with the fact that the commutator of
covariant derivatives is the magnetic field.
The commutators of ${\hat R}_{+i}$ and ${\hat R}_{-i}$ are in the
Lie algebra of $U(k)$; in the case of ${\mathbb {CP}}^k$ these correspond to constant magnetic fields. In particular we can specify the background field by specifying the right action of $U(k)$ on the wavefunctions. 
\beqar
{\hat R}_a ~\Psi^J_{m; \alpha} (g) &=&
(T_a^{\tilde{J}})_{\alpha \beta} \Psi^J_{m; \beta} (g) \label{right1}\\
{\hat R}_{k^2 +2k} ~\Psi^J_{m; \alpha} (g) &=& - {n k\over \sqrt{2 k
(k+1)}}~\Psi^J_{m; \alpha} (g) \label{right2}
\eeqar
where the index $m=1,\cdots, {\rm dim}J$ represents the state within the $SU(k+1)$ representation $J$ and therefore counts the degeneracy of the Landau level. The first of these equations shows that the wavefunctions $\Psi^J_{m; \alpha}$ transform, under right
rotations, as a  representation $\tilde{J}$ 
of
$SU(k)$. 
$(T^{{\tilde J}}_a)_{\alpha \beta}$ are the representation matrices for the
generators of
$SU(k)$ in the representation ${\tilde J}$ and 
$n$ is an integer characterizing the abelian part of the background field.
$\alpha ,\beta$ label states within the $SU(k)$ representation ${\tilde J}$
(which is itself
contained in the representation $J$ of $SU(k+1)$). The index $\alpha$ in the
wavefunctions  $\Psi^J_{m; \alpha} (g)$ characterizes the nonabelian charge of the underlying fermion fields. 

 In terms of $\D$-functions, the correctly normalized wavefunctions are given by
 \beq
\Psi^J_{m; \alpha} (g) = \sqrt{N} \, \bra{J, m} g \ket{J, \alpha, n} = \sqrt{N} ~{\cal D}^J_{m; \alpha}(g)
\label{normalization}
\eeq
where $N= {\rm dim} J$ and the following orthogonality theorem has been used 
\beq
\int d\mu (g) ~\D^{*J}_{m;\alpha} (g)~\D^{J}_{m';\alpha '}
(g) ~=~ {\delta_{mm'}\delta_{\alpha \alpha '}\over N}
\label{orthogonality}
\eeq
$d\mu (g)$ is the Haar measure on $SU(k+1)$ normalized to
unity.

In the absence of a confining potential, the Hamiltonian $H$ for the Landau problem is proportional to the covariant Laplacian on 
${\mathbb{CP}}^k$, namely
\beq
H \, \Psi = - {1\over 4 m} (\D_{+i} \D_{-i} + \D_{-i} \D_{+i} ) \, \Psi
\label{hamiltonian}
\eeq
which apart from additive constants can be reduced to the form $\sum_{i} {\hat R}_{+i} {\hat R}_{-i}$. Thus the lowest Landau level wavefunctions satisfy the holomorphicity condition
\beq
\hat{R}_{-i} \, \Psi = 0
\label{holo}
\eeq
The conditions (\ref{right1}), (\ref{right2}) and (\ref{holo}) completely fix the representation $J$ and therefore the degeneracy of the lowest Landau level. 

First we consider the lowest Landau wavefunctions for the case of an abelian background magnetic field. In that case the state $\ket{J, n}$ corresponds to the singlet representation of $SU(k) \in SU(k+1)$ with a $U(1)$ charge proportional to $n$ as specified in (\ref{right2}), namely $R_3=-n/2$.  These can be thought of as the coherent states for ${\mathbb {CP}}^k$, written explicitly in terms of complex coordinates,
\beqar
\Psi_{i_1 i_2 \cdots i_k}&=& \sqrt{N} \left[ {n! \over {i_1! i_2! ...i_k!
(n-s)!}}\right]^{\half} ~ {z_1^{i_1} z_2^{i_2}\cdots z_k^{i_k}\over
(1+\bz \cdot z )^{n \over 2}}~,~~~~~~~~~~~~  \nonumber\\
s &=& i_1 +i_2 + \cdots +i_k ~,~~~0\le i_i \le n~~, ~~~  0 \le s \le n \label{wav}
\eeqar
These wavefunctions form a symmetric, rank $n$ representation $J$ of $SU(k+1)$. The dimension of this representation, which is also the LLL degeneracy, is 
\beq
N={\rm dim} J = {{(n+k)!} \over {n! k!}} 
\label{dim}
\eeq
The volume element for ${\mathbb{CP}^k}$ is 
\beq
d\mu = {k! \over \pi^k} {d^2z_1 \cdots d^2z_k
\over (1+ \bz \cdot z)^{k+1}}
\label{vN40}
\eeq
We have chosen the normalization such that the total volume, $\int d\mu$, is $1$.

In the case of a $U(1) \times SU(k)$ nonabelian background,
it is convenient to label the irreducible representation of $SU(k+1)_R$ by $(p+l,
q+l')$ corresponding to the tensor \cite{KN1}
\beq
{\cal T}^{a_1...a_q \g_{1}...\g_{l'}}_{b_1...b_p \d_{1}...\d_{l}} \equiv
{\cal T}^{q,l'}_{p,l} \label{7a}
\eeq
where $p,q$ indicate $U(1)$ indices and $l,l'$ indicate $SU(k)$ indices, namely
$a$'s and $b$'s take the value $(k+1)$ and $\gamma$'s and $\delta$'s take
values $1,\cdots,k$. 

The right hypercharge corresponding to (\ref{right2}) is
\beq
\sqrt{2k(k+1)} R_{k^2+2k} = -k(p-q)+l-l' = -nk
\label{7b}
\eeq
The fact that $n$ has to be integer implies that $(l-l')/k$ is an integer,
thus constraining the possible $SU(k)_R$ representations $\tilde{J}$.

Further, as explained in detail in \cite{KN1}, the lowest Landau level states correspond to $q=0,~l=0$. So the LLL states 
we consider correspond to the tensor ${\cal T}^{l'}_p$, where
$p=n-{l' \over k}$ and $l'=jk,~j=1,2,\cdots$.

\section{Entanglement Entropy for $\nu=1$}

The entanglement entropy $S$ for the $\nu =1$ lowest Landau level quantum Hall states is given by
\beq
S = - \sum_{m=1}^{N} \left[ \lambda_m \log \lambda_m + (1 - \lambda_m) \log (1-\lambda_m)\right]
\label{entropy}
\eeq
where the index $m$ counts the degeneracy and $\lambda$'s are the eigenvalues of the two-point correlator $C(z, z')$ \cite{Sierra}, 
\beq
C(z, z') = \sum_{m=1}^{N} \Psi_m^{*} (z) ~\Psi_m (z')
\eeq
where $z,z'$ are restricted to be inside the domain $D$. We choose $D$ to be the spherically symmetric region of ${\mathbb {CP}}^k$ satisfying $z'\cdot \bar{z} \le R^2$. For ${\mathbb {CP}}^1 \sim S^2$, this region is a polar cap bounded by a latitude at $\theta$, with $R = \tan \theta/2$ via stereographic projection.

The diagonalization of $C(z, z')$ gives the result
\beqar
\int C(z, z') \Psi^*_l (z') d\mu (z') & = & \sum_{m=0}^{N} \Psi_m^{*} (z)~\int \Psi_m (z') \Psi^*_l (z') d\mu (z') \nonumber \\
&=& \lambda _{l} \Psi_l^{*} (z) 
\label{diagonal}
\eeqar
where
\beq
\lambda_l = \int_{D} |\Psi_{l}|^2 d \mu
\label{lambda}
\eeq
The second line in (\ref{diagonal}) is due to the fact that the angular integration over the spherically symmetric region $D$ will give $\int_{D} \Psi_m (z') ~\Psi_l^*(z') d \mu (z') = \delta_{lm}~\lambda_l$.

We now proceed to calculate the eigenvalues $\lambda$ and subsequently the entanglement entropy for the case of an abelian and nonabelian magnetic field backgrounds. 

\subsection{ ${\mathbb {CP}}^k$ and abelian magnetic field background} 

The lowest Landau level wavefunctions for ${\mathbb {CP}}^k$ in the case of an abelian background magnetic field are given in (\ref{wav}). The corresponding eigenvalues of the two-point correlator are 
\beqar
\lambda_{i_1 i_2 \cdots i_k}  &=& \int_D d\mu \,\Psi^*_{i_1 i_2 \cdots i_k} (r) \Psi_{i_1 i_2 \cdots i_k} (r)
\nonumber\\
&=& {{(n+k)! \over {{i_1}! {i_2}!\cdots {i_k}! (n-s)!}}} ~
\int_D  {{(\bar{z}_1 z_1)^{i_1} (\bar{z}_2 z_2)^{i_2} \cdots (\bar{z}_k z_k)^{i_k}} \over {(1+\bar{z}\cdot z)^{n+k+1}}}  {d^2z_1 \cdots d^2z_k
\over {\pi^k}}
\label{lambda1}
\eeqar
where $s= i_1 +i_2 + \cdots +i_k$. We perform the angular integration using the parametrization $z_i = x_{2i-1} + i x_{2i}$, where
\beqar
x_1 &=& \rho \cos (\phi_1) \nonumber \\
x_2 &=& \rho \sin (\phi_1) \cos (\phi_2) \nonumber \\
\cdots  \nonumber \\
x_{2k-1} &=& \rho \sin (\phi_1) \sin (\phi_2) \cdots \sin (\phi_{2k-2}) \cos (\phi_{2k-1}) \nonumber \\
x_{2k} &=& \rho \sin (\phi_1) \sin (\phi_2) \cdots \sin (\phi_{2k-1}) 
\label{para}
\eeqar
and $0 \le \phi_1, \phi_2, \cdots , \phi_{2k-2} \le \pi~~,~~0 \le \phi_{2k-1} \le 2 \pi$. 
Using the fact that in terms of this parametrization
\beq
d^2z_1 \cdots d^2z_k = \rho^{2k-1} d \rho ~\sin( \phi_1)^{2k-2} ~d\phi_1~ \sin(\phi_2)^{2k-3}  ~d\phi_2 \cdots \sin (\phi_{2k-2}) d\phi_{2k-2} d\phi_{2k-1}
\label{22}
\eeq
and 
\beq
\int_{0}^{\pi} (\sin \phi)^{2i} ~d \phi = \sqrt{\pi} ~{ \Gamma (i + \half) \over \Gamma (i + 1)} 
\label{23}
\eeq 
we find, after doing the angular integrations, that
\beq
\lambda_{i_1 i_2 \cdots i_k} \equiv \lambda_s = {(n+k)! \over {(n-s)! (s+k-1)!}} \int_0^{R^2} {x^{s+k-1} \over {(1+x)^{n+k+1}}} dx
\label{24}
\eeq
For each value of $s= i_1 +i_2 + \cdots +i_k$, the eigenvalue $\lambda_s$ has a degeneracy $d_s = {(s+k-1)! / {s! (k-1)!}}$. 

The expression for the entanglement entropy is 
\beq
S=  \sum_{s=0}^{n} ~{(s+k-1)! \over {s! (k-1)!}}  \left[ -\lambda_s \log \lambda_s - (1 - \lambda_s) \log (1-\lambda_s)\right]
\label{25}
\eeq

We will now evaluate the entanglement entropy using a semiclassical approximation and relate this to the area of the region $D$. This is possible when the $U(1)$ charge $n$, which controls the dimensionality of the lowest Landau Hilbert space, becomes very large. 

Making a change of variables to $t = x/(1+ x)$, the expression for the eigenvalues $\lambda$'s in (\ref{24}) can be written as,
 \beqar
 \lambda_s &=& {{ (n+k)!} \over {(s+k-1)! (n-s)!}}
 \int_0^{t_0} dt \, t^{s+k-1} (1-t)^{n-s} \nonumber\\
 &=& {{ (n+k)!} \over {(s+k-1)! (n-s)!}} B(t_0; s+k, n-s+1)
 \label{beta}
\eeqar
where $t_0 = R^2/(1+R^2)$ and $B(z; m_1,m_2)$ is the incomplete beta function. For large $n$ this is amenable to a semiclassical calculation as shown in \cite{VPN}. We will follow that derivation here. Eq. (\ref{beta}) can be written as 
\beqar
\lambda_s &=& { (n+k)! \over {(s+k-1)! (n-s)!}} \int_0^{t_0} \, dt\, e^{F(t)} \nonumber  \\
 F(t)&=&(s+k-1) \log t + (n-s) \log(1-t) 
 \label{49}
\eeqar
The maximum of $F(t)$ occurs at $t^* = s+k-1/(n+k-1)$. Expanding $F(t)$ around $t^*$ we find that $e^{F}$ becomes a Gaussian function centered around $t^*$. In fact,
\beq
{d^2 F \over {dt^2}}\big\vert_{t^{*} }= - {(n+k-1)^3 \over {(n-s) (s+k-1)}} 
\eeq 
which implies that the width of the Gaussian is very narrow for all $s$. For small $s$ the center of the Gaussian, $t^* \sim 0$, falls within the range of integration and we find that $\lambda_s \sim 1$. For large $s \sim n$ the center of the Gaussian, $t^* \sim 1  > t_0$, falls outside the range of integration and therefore $\lambda_s \sim 0$. The middle of the transition occurs at $s^*$ such that $t^* = t_0$, namely
\beqar
t^*= {{s^*+k-1} \over {n+k-1}} = t_0 ~~~~\Rightarrow && s^* = t_0(n+k-1) - (k-1) \\
&&n-s^* = (n+k-1)(1-t_0) \nonumber
\label{t*}
\eeqar
Expanding $F(t)$ around $t_0$ in (\ref{t*}) we find
\beq
F(t) = (n+k -1) \left[ t_0 \log t_0 + (1-t_0) \log (1-t_0) \right]
-   {(n+k-1)\over 2 t_0 (1-t_0) }  (t-t_0)^2 + \cdots 
\label{52}
\eeq
Using this expression we find that for large $n$
\beq
\int_0^{t_0} e^{F(t)} \sim e^{F(t_0)} ~ \int_0^{t_0} \exp \left[ - {(n+k-1)\over 2 t_0 (1-t_0) }  (t-t_0)^2 \right] = e^{F(t_0)} \sqrt{ {{\pi t_0 (1-t_0)} \over {2(n+k-1)}}}
\label{53}
\eeq 
Substituting this in (\ref{49}) and using Stirling's formula $n! = \sqrt{2 \pi n} ~(n/e)^n$, we find that 
\beq
\lambda_{s_*} \approx { n+k \over {2(n+k-1)}} \rightarrow {1 \over 2} 
\label{54}
\eeq
{\it {for any $t_0$}}. The value of $t_0$ is controlled by $R$, which characterizes the size of the spherical domain $D$ and the above calculation shows that $\lambda_s$ is significantly different from 0 or 1 only for $s$ such that the corresponding wavefunctions are localized very near the boundary of the entangling surface.

For large $n$ we can define a variable $y = s/(n+k-1)$, ~$0 \le y \le 1$, and consider $\lambda$ as a continuous function of $y$. From what we have seen before $\lambda \rightarrow 1$ as $y \rightarrow 0$, $\lambda \rightarrow 0$ as $y \rightarrow 1$ and $\lambda \rightarrow 1/2$ as $y \rightarrow s^* /(n+k-1)=t_0$. In deriving a semiclassical expression for the entanglement entropy we will also need to calculate the derivative of $\lambda$ at the transition region, namely ${d \lambda \over dy}|_{y =t_0}$. For that we have to calculate the difference $\lambda_{s^* +i} - \lambda_{s^*}$. For $s=s^* +i$, the maximum of $F(t)$ occurs at 
\beq
t_1 = {s_* +k -1 \over n+k-1} + {i \over n+k-1} =  t_0 +  \epsilon, 
\label{55}
\eeq
where $\epsilon = i / (n+k-1) \ll 1$ for small $i$ and large $n$. We now expand $F$ in (\ref{49}) around $t_1$, but because the peak has been shifted beyond the upper limit of integration, (\ref{53}) will give an extra contribution proportional to $\epsilon$ for small $\epsilon$,
\beq
\int_0^{t_0} \exp \left[ - {(n+k-1)\over 2 t_1 (1-t_1) }  (t-t_1)^2 \right] \sim \sqrt{ {{\pi t_0 (1-t_0)} \over {2(n+k-1)}}} - \epsilon
\label{56}
\eeq
Using Stirling's formula and taking $\epsilon \rightarrow 0$ we find 
\beq
{d\lambda \over dy}\big\vert_{y=t_0} = lim_{\epsilon \rightarrow 0} {{\lambda_{s^* + i} - \lambda_{s^*}} \over \epsilon} \sim - \sqrt{n+k-1 \over {2 \pi t_0 (1-t_0)}} + {\cal{O}} ( {1 \over \sqrt{n}}) 
\label{57}
\eeq
Figures 1 and 2 show plots of $\lambda_s$ for different values of $k$ and $t_0=R^2/(1+R^2)$. 

\begin{figure}[h]
\begin{center}
\begin{minipage}{5cm}
\scalebox{.5}{\includegraphics{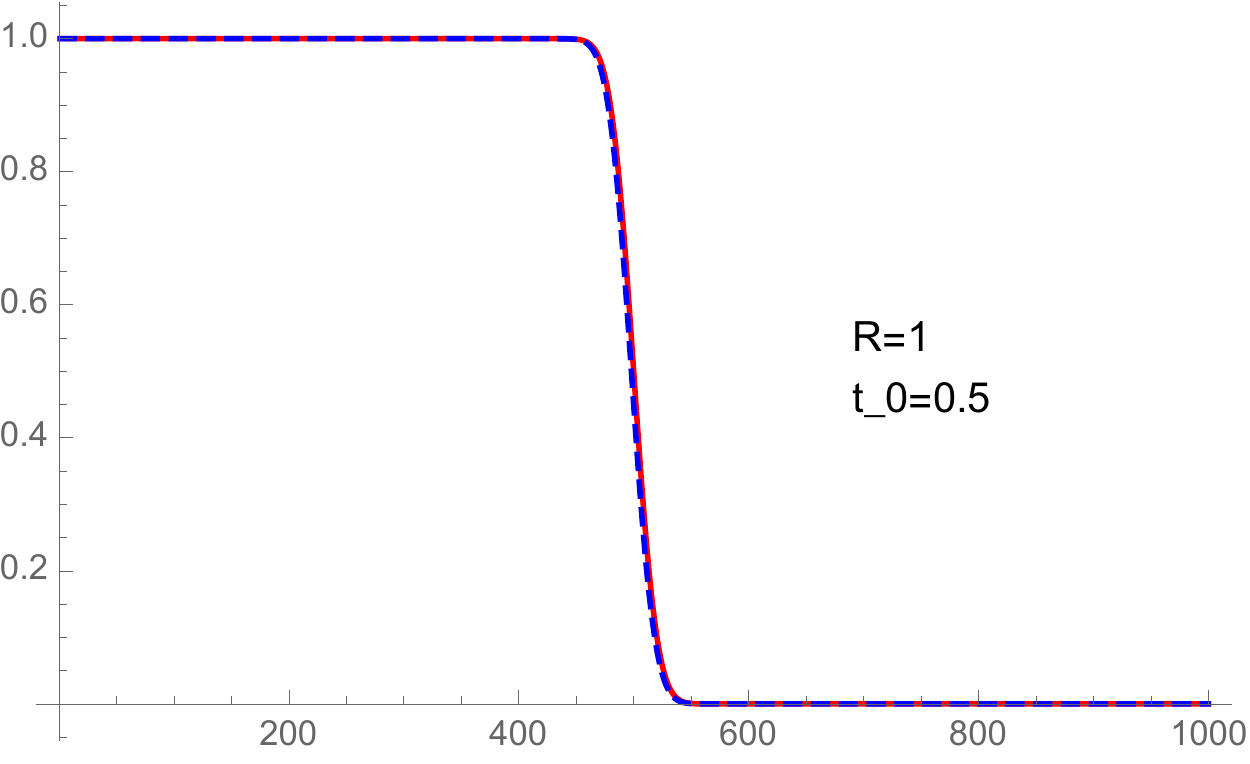}}
\caption{Plots of $\lambda_s$ as a function of $s$ for $k=1$ (red), $k=5$ (blue dashed) and $n = 1000$ and $R=1$}
\label{g1}
\end{minipage}
\hskip 1in
\begin{minipage}{5cm}
\scalebox{.5}{\includegraphics{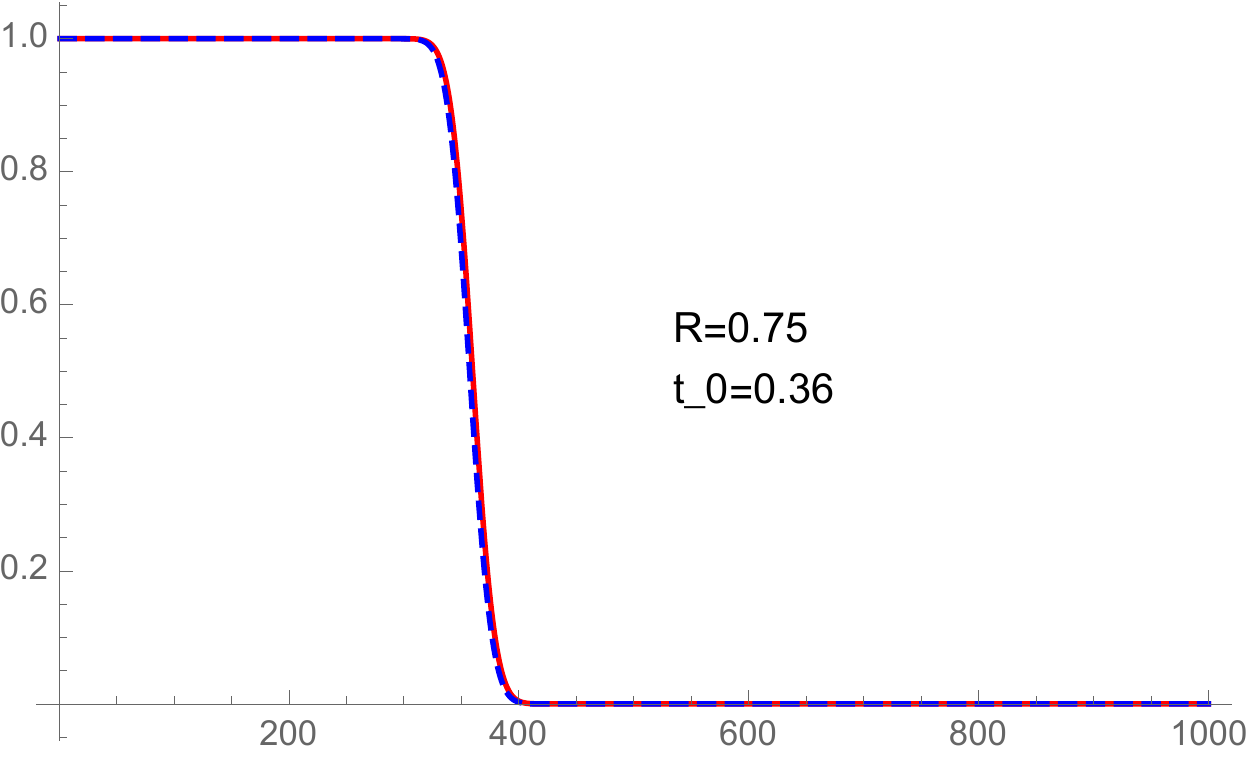} }
\caption{Plots of $\lambda_s$ as a functio of $s$ for $k=1$ (red), $k=5$ (blue dashed) and $n = 1000$ and $R=0.75$}
\label{g2}
\end{minipage}
\end{center}
\end{figure}

We found in (\ref{25}) that the expression for the entanglement entropy is 
\beqar
S &= &\sum_{s=0}^{n} ~{(s+k-1)! \over {s! (k-1)!}}  ~H_s \nonumber \\
H_s & = & -\lambda_s \log \lambda_s - (1 - \lambda_s) \log (1-\lambda_s)
\label{58}
\eeqar
It is clear from the Figures 1 and 2 that $H_s$ is nonzero only for values of $s$ very near the transition region where  $\lambda_{s^*} = 1/2$. We can then expand $H(\lambda(y))$ around the value $\lambda=1/2$,
\beq
H(\lambda(y)) = H(t_0) + {1 \over 2} {d^2 H \over {dy^2}}(y-t_0)^2 + \cdots
\label{59}
\eeq
where
\beqar
{dH \over dy}|_{y=t_0} & =& {dH \over d\lambda} {d\lambda \over dy}|_{\lambda=1/2} =0   \nonumber \\
{ d^2 H \over {d^2 y}}|_{y=t_0} & =&  {dH \over d\lambda} {d^2 \lambda \over dy^2}  + {d^2 H \over d\lambda^2}({d\lambda \over dy})^2  |_{\lambda=1/2} =~-~{4(n+k-1) \over 2\pi t_0 (1-t_0)}
\label{60}
\eeqar
Since $H$ has a narrow support around $\lambda = 1/2$ it can be approximated by the Gaussian 
\beqar
H(y) & = & H_0 ~ \exp \bigl[{1 \over 2}{d^2 H \over d\lambda^2}({d\lambda \over dy})^2 (y-t_0)^2\bigr] \nonumber \\
&=& \ln 2 ~e^{-{(n+k-1) \over{ \pi \ln 2 t_0 (1 -t_0)}} (y-t_0)^2} 
\label{61}
\eeqar
We can rescale to $s=y(n+k-1)$ and $s^* = t_0(n+k-1)-(k-1)$ to obtain the semiclassical Gaussian approximation to $H_s$ as 
\beq
H_{s,k} = \ln 2~ \exp \bigl[ - {1 \over {\pi \ln(2) t_0 (1-t_0)}} {(s+k-1-t_0 (n+k-1))^2 \over {n+k-1}} \bigr]
\label{62}
\eeq
Figures 3 and 4 show the validity of the Gaussian approximation by comparing (\ref{62}) to the exact expression (\ref{58}), (\ref{beta}).

\begin{figure}[h]
\begin{center}
\begin{minipage}{5cm}
\scalebox{.5}{\includegraphics{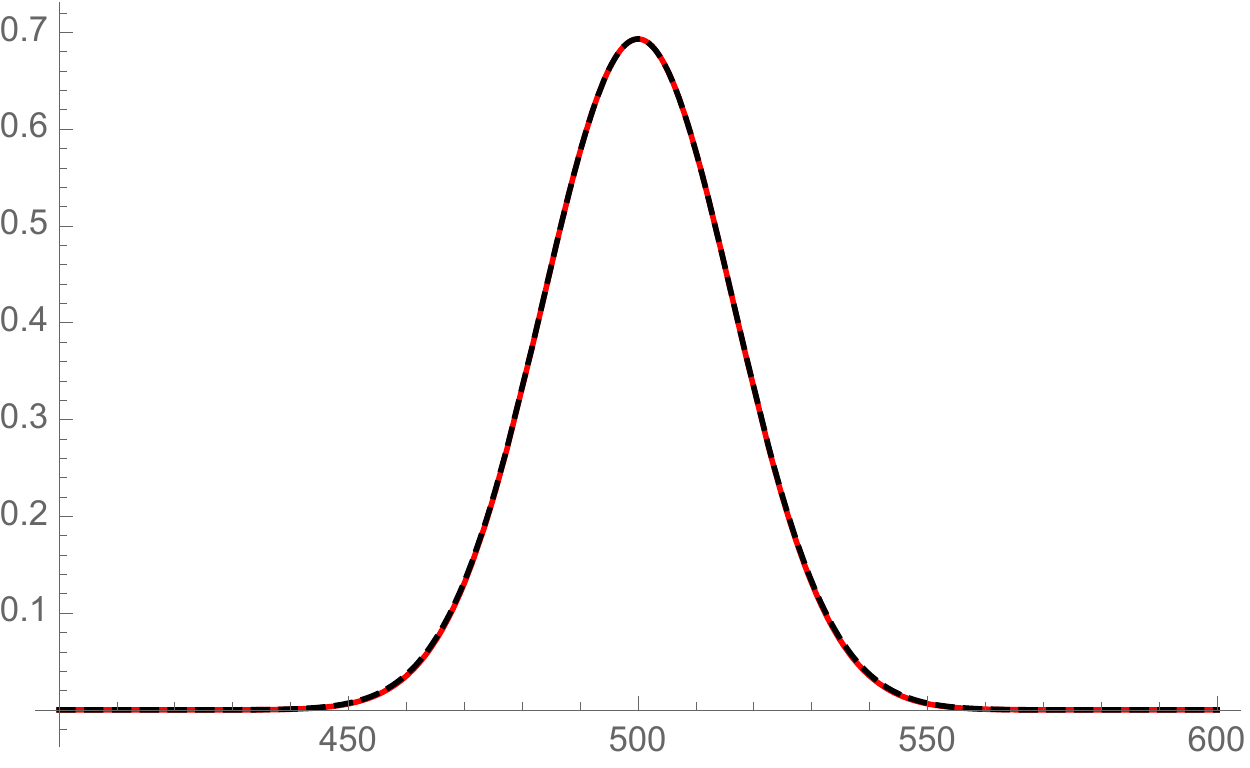}}
\caption{Plots of $H_s$ exact (red) and Gaussian approximation (blue dashed) as a function of $s$ for $k=1$ and $n = 1000$ and $R=1$}
\label{g1}
\end{minipage}
\hskip 1in
\begin{minipage}{5cm}
\scalebox{.5}{\includegraphics{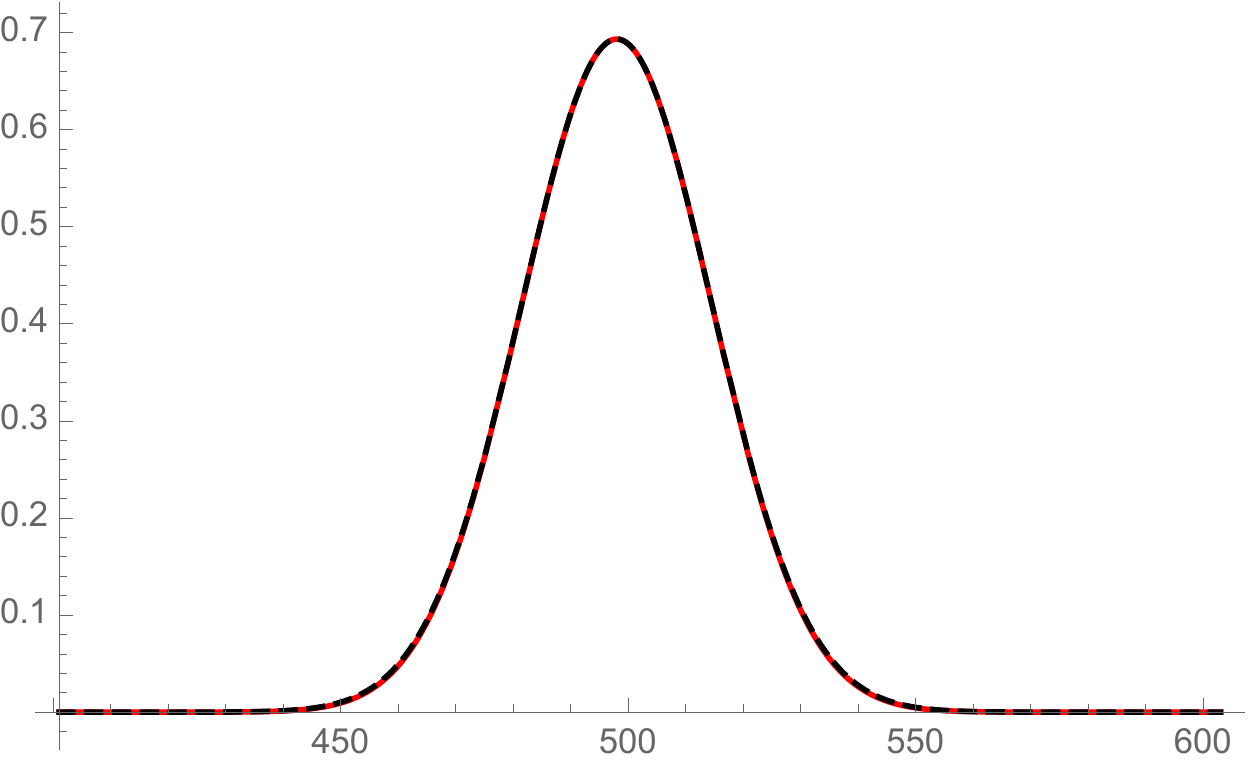} }
\caption{Plots of $H_s$ exact (red) and Gaussian approximation (blue dashed) as a function of $s$ for $k=5$ and $n = 1000$ and $R=1$}
\label{g2}
\end{minipage}
\end{center}
\end{figure}
We can now use (\ref{62}) to analytically calculate the entropy in (\ref{58}) for large $n$ by converting the sum into an integral over the variable $y$ 
\beqar
S & \sim & n {s^{*k-1} \over (k-1)!} \ln 2 \int_0^1 \exp \bigl[ - {(n+k-1) \over { \pi ~\ln 2~t_0 (1-t_0)}} (y-t_0)^2 \bigr] ~dy  \nonumber \\
& \sim & n^{k - \half} {t_0^{k-1} \over (k-1)!} ~\pi ~( \ln 2)^{3/2} \sqrt{ t_0 (1-t_0)} \nonumber \\
& \sim & n^{k - \half} ~{\pi ~( \ln 2)^{3/2}  \over (k-1)!} ~{R^{2k-1} \over {(1+R^2)^k}} 
\label{60a}
\eeqar

The fact that the entropy is proportional to the entangling area $R^{2k-1} \over {(1+R^2)^k}$ has to do with the fact that only wavefunctions localized around the entanglement  boundary with corresponding eigenvalues $\lambda \sim 1/2$ contribute to the entropy.

For $k=1$ this agrees with the result found in \cite{Sierra}. In the case of the QHE on the sphere the entangling surface is a circle of perimeter $L= 2 \pi \sin\theta$, where, based on the stereographic projection
$2R /(1+R^2)  = \sin\theta$. Scaling the radius of the entangling surface by $\sqrt{n/2}$ (for QHE on $S^2$ the monopole charge, magnetic field and radius of the sphere are related by $n=2B r^2$) we find the area law quoted in \cite{Sierra}
\beq
S(k=1) = {{ \sqrt(2) (\ln 2)^{3/2}} \over 4} ~L  = 0.204~ L
\label{64}
\eeq

The normalized volume element (\ref{vN40}), upon angular integration can be written in terms of the radial variable $\rho$ defined in (\ref{para}) 
\beq
d\mu = {k! \over \pi^k} {d^2z_1 \cdots d^2z_k
\over (1+ \bz \cdot z)^{k+1}} = 2k { \rho^{2k-1} \over {(1 + \rho^2)^k}} ~{ d \rho \over {1 + \rho^2}} 
\label{61}
\eeq 
where $e_{\rho} = d \rho / (1 + \rho^2) $ is the vierbein along the radial direction $\rho$. This defines the geometric area of the entangling surface (with volume normalized to 1) to be $A_{\rm geom} =   2k { R^{2k-1} \over {(1 + R^2)^k}} $. On the other hand the phase-space volume which is proportional to the degrees of freedom is $V_{\rm phase~space} = { n^k \over k!} \int d \mu$. This then defines a phase-space surface area
\beq
A_{\rm phase~space} = {n^{k-\half}  \over k!} A_{\rm geom} = 2~{n^{k - \half}   \over (k-1)!} ~{R^{2k-1} \over {(1+R^2)^k}} 
\label{62a}
\eeq
Scaling the entanglement entropy in (\ref{60a}) in terms of this phase-space area we derive a universal expression valid in all dimensions, with a proportionality constant independent of $k$, namely
\beq
S \sim {\pi \over 2} ( \ln 2)^{3/2} ~A_{\rm phase~space}
\label{63}
\eeq

\subsection{ ${\mathbb {CP}}^2$ and nonabelian magnetic field background} 

The derivation of the entanglement entropy in the case of a nonabelian background magnetic field is more involved. As mentioned in section 2 the LLL states form irreducible representations of $SU(k+1)$ of the form ${\cal T}^{l'}_p$, where
$p=n-{l' \over k}$ and $l'=jk,~j=1,2,\cdots$. We will elucidate the calculation of the entanglement entropy for the special case of ${\mathbb {CP}}^2$ with a nonabelian magnetic field $U(1) \times SU(2)$ for the lowest value of $l'$ , namely $l'=k=2$ and $p=n-1$. The derivation for other values of $k$ and $l'$ follows similar ideas. The dimension of this representation and therefore the degeneracy of the corresponding LLL is \cite{KN1}
\beq
N = {3 n (n+3) \over 2} 
\label{26}
\eeq
In identifying the corresponding wavefunctions we consider the states $\bra{ m}  \hat{g } \ket{w} $, where the states on the right are of the form ${\cal T}^2_p$ with two up indices and transforming as the $\tilde{J}=1$ triplet representation of $SU(2) \in SU(3)$. (Since the lowest allowed value for $l'$ is 2, based on (\ref{7b}) and following comments, the doublet representation is not allowed for ${\mathbb {CP}}^2$.) The corresponding group elements in the appropriate representation can be constructed in terms of products of elements of the $3 \times 3$ matrix $g$ which forms the fundamental representation of $SU(3)$ and its conjugate $g^*$. We need $p$ copies of $g$ and two copies of $g^*$ to match the structure of the ${\cal T}^2_p$ representation. In terms of these matrices, choosing the state $\ket{w}$ as explained above, we get
\beq
\bra{i_1 i_2 ; j_1 \cdots j_p} \hat{g} \ket{3 \cdots 3; \alpha \beta} \sim g^{*i_1\alpha}g^{*i_2\beta}g_{j_1 3} \cdots g_{j_p 3}
\label{general}
\eeq
where $i,j =1,2,3$ and $\alpha, \beta = 1,2$.
Within (\ref{general}) there are three distinct series, each one forming an $SU(2)$ multiplet under the left transformations, and all of them together comprising the full $\nu=1$ lowest Landau level $SU(3)$ representation. The three such series are of the form:

\underline {Series 1}

\beq
\Psi^{(1)}_{(\alpha\beta)} \sim g^{*3\alpha}~g^{*3\beta} ~ (g_{13})^{l}~(g_{23})^{m-l}~(g_{33})^{n-1-m}
\label{27}
\eeq
where $l = 0,\cdots,m$ and $m=0,\cdots,n-1$. For each $m$, (\ref{27}) form an $SU(2)$ left representation with $j= m/2$. There are $\sum_0^{n-1} (m+1) = n(n+1) /2$ such states. 

\underline {Series 2}

\beq
\Psi^{(2)}_{(\alpha\beta)} \sim ( g^{*\gamma\alpha}~g^{*3\beta} +g^{*3\alpha}~g^{*\gamma\beta}) ~ (g_{13})^{l}~(g_{23})^{m-l}~(g_{33})^{n-1-m}
\label{28}
\eeq
For each $m$, (\ref{28}) form an $SU(2)$ left representation with $j= (m+1)/2$. There are $\sum_0^{n-1} (m+2)= n(n+3) /2$ such states. 

\underline {Series 3}

These are of the form
\beq
\Psi^{(3)}_{(\alpha\beta)} \sim  ( g^{*\gamma\alpha}~g^{*\delta\beta} +  g^{*\gamma\beta}~g^{*\delta\alpha})~ (g_{13})^{l}~(g_{23})^{m-l}~(g_{33})^{n-1-m}
\label{29}
\eeq
For each $m$, (\ref{29}) form an $SU(2)$ left representation with $j= (m/2)+1$. There are $\sum_0^{n-1} (m+3)= n(n+5) /2$ such states. 

Considering all three series together, the total number of states are $N = 3n(n+3) /2$ confirming the result in (\ref{26}). We now proceed to normalize the above wavefunctions. In doing so we will use the fact that the elements $g_{i 3}$ can be written in terms of the complex coordinates parametrizing ${\mathbb {CP}}^2$, namely
\beqar
 g_{\alpha 3}& = & {z_{\alpha} \over {\sqrt{1 + \bz \cdot z}}}~,~~~~~~~\alpha=1,2 \nonumber \\
 g_{3 3} & = & {1 \over {\sqrt{1 + \bz \cdot z}}} 
\label{31a}
\eeqar
where $\bz \cdot z = \bz_1 z_1 + \bz_1 z_1$. 

States, with the correct normalization within each series, can be explicitly constructed by starting with the highest weight state and applying the lowering operator $J_{-}$ as follows
\beqar
J_{-} ~g_{13} = g_{23} ~~&,& ~~ J_{-} ~g_{23} =0 \nonumber \\
J_ {-}~ g^{*2i} = - g^{*1i} ~~&,&~~J_ {-}~g^{*1i} = 0 
\label{30}
\eeqar

\underline {Series 1 normalization}

The highest weight state within this $SU(2)$ multiplet is the state of the form
\beq
\ket{J,J} = C_1 ~g^{*3\alpha}~g^{*3\beta}  ~ (g_{13})^{m}~(g_{33})^{n-1-m}
\label{31}
\eeq
where $C_1$ is the normalization factor to be determined. The rest of the states are obtained by applying the lowering operator $J_{-}$ whose action is indicated in (\ref{30}), namely
\beq
\ket{J, J-l}  =  C_1 ~ g^{*3\alpha}~g^{*3\beta}  ~ (g_{13})^{m-l}~(g_{23})^{l}~(g_{33})^{n-1-m}
\label{32}
\eeq
where $l=0,1,\cdots, m$. 
Using (\ref{31}) and the fact that $g^{\dagger} g = 1$ we find that
\beq
\sum _{\alpha\beta} \Psi_{\alpha\beta}^{*(1)}\Psi_{\alpha\beta}^{(1)} = |C_1|^2 \left[ 1 - { 1 \over {1 + \bz \cdot z}} \right] ^2 {{ (\bz_1 z_1)^{m-l} (\bz_2 z_2)^{l}} \over {(1 + \bz \cdot z)^{n-1}}}
\label{32a}
\eeq 

The ${\mathbb {CP}}^2$ volume element is $d\mu = {2 \over \pi^2} {d^2z_1 d^2z_2
\over (1+ \bz \cdot z)^{3}}$. Using the relation
\beq
\int  {{ (\bz_1 z_1)^l (\bz_2 z_2)^{m}} \over {(1 + \bz \cdot z)^{n+1}}} ~d \mu = 2 {l! m! (n+1-l-m)! \over (n+3)!}
\label{33}
\eeq
we find that the correctly normalized wavefunction is of the form
\beq
\Psi^{(1)}_{(l,m;\alpha\beta)} = \sqrt{{(n+3)! \over {2~ l! (m-l)! (m+2)(m+3) (n-1-m)!}}} ~g^{*3\alpha}~g^{*3\beta} ~ (g_{13})^{m-l}~(g_{23})^{l}~(g_{33})^{n-1-m}
\label{34}
\eeq
where $l=0,1,\cdots, m$. 

\underline {Series 2 normalization}

The highest weight state within this $SU(2)$ multiplet is given by 
\beq
\ket{J,J} = C_2 ( g^{*2\alpha}~g^{*3\beta} +g^{*3\alpha}~g^{*2\beta}) ~ (g_{13})^{m}~(g_{33})^{n-1-m}
\label{35}
\eeq
Acting with the lowering operator $J_{-}$ as before we obtain the rest of the states which are of the form
\beqar
\ket{J, J-l} & = & C_2 \Big[ -l {m! \over (m-l+1)!} (g^{*1\alpha}~g^{*3\beta} +g^{*3\alpha}~g^{*1\beta}) ~ (g_{13})^{m-l+1}~(g_{23})^{l-1} \nonumber \\
& + &{m! \over (m-l)!} (g^{*2\alpha}~g^{*3\beta} +g^{*3\alpha}~g^{*2\beta}) ~ (g_{13})^{m-l}~(g_{23})^{l}  \Big] (g_{33})^{n-1-m}
\label{36}
\eeqar
where $l=0,1,\cdots, m+1$. Using again the relation (\ref{33}) we find the normalized wavefunctions to be of the form
\beqar
\Psi^{(2)}_{(l,m;\alpha\beta)} &=& \sqrt{{(n+3)! \over {4 ~l! (m-l+1)! (m+1)(m+3)(n+1) (n-1-m)!}}} \nonumber \\
&\times & \Big[ -l  (g^{*1\alpha}~g^{*3\beta} +g^{*3\alpha}~g^{*1\beta}) ~ (g_{13})^{m-l+1}~(g_{23})^{l-1} \nonumber \\
  &+& (m-l+1) (g^{*2\alpha}~g^{*3\beta} +g^{*3\alpha}~g^{*2\beta}) ~ (g_{13})^{m-l}~(g_{23})^{l}  \Big] (g_{33})^{n-1-m}
\label{37}
\eeqar
$l=0,1,\cdots, m+1$. 

\underline {Series 3 normalization}

The highest weight state in series 3 is given by
\beq
\ket{J,J} = C_3 ~ g^{*2\alpha}~g^{*2\beta}  ~ (g_{13})^{m}~(g_{33})^{n-1-m}
\label{38}
\eeq
Every other state in this multiplet is constructed as before by applying $l$ times the lower operator $J_{-}$,  producing
\beqar
\ket{J, J-l} & = & C_3 \Big[{m! \over (m-l)!} g^{*2\alpha}~g^{*2\beta}  ~ (g_{23})^l~(g_{13})^{m-l} \nonumber \\
&-&l {m! \over (m-l+1)!} (g^{*1\alpha}~g^{*2\beta} +g^{*2\alpha}~g^{*1\beta}) ~ (g_{13})^{m-l+1}~(g_{23})^{l-1} \nonumber \\
& + &l(l-1){m! \over (m-l+2)!} g^{*1\alpha} ~ (g_{13})^{m-l+2}~(g_{23})^{l-2}  \Big] (g_{33})^{n-1-m}
\label{39}
\eeqar 
where $l=0,1,\cdots, m+2$.  Using (\ref{33}) we find the normalized wavefunctions to be of the form
\beqar
\Psi^{(3)}_{(l,m;\alpha\beta)} &=& \sqrt{{(n+3)! \over {2 ~l! (m-l+2)! (m+1)(m+2)(n+1) (n+2) (n-1-m)!}}} \nonumber \\
&\times & \Big[ l (l-1) g^{*1\alpha}~g^{*1\beta} ~ (g_{13})^{m-l+2}~(g_{23})^{l-2} \nonumber \\
  &-& l (m-l+2) (g^{*1\alpha}~g^{*2\beta} +g^{*2\alpha}~g^{*1\beta}) ~ (g_{13})^{m-l+1}~(g_{23})^{l-1} \nonumber \\
  &+& (m-l+1)(m-l+2)  g^{*2\alpha}~g^{*2\beta} ~ (g_{13})^{m-l}~(g_{23})^{l} \Big] (g_{33})^{n-1-m}
\label{40}
\eeqar
$l=0,1,\cdots, m+2$. 

The two-point correlator carries nonabelian indices and is defined as 
\beq
C_{ab}(r,r') = \sum_{A} \Psi^*_{A;a}(r)~\Psi_{A;b}(r')
\label{cab}
\eeq 
where we have denoted collectively the left indices by $A=(l,m)$ and the right (nonabelian) indices by $a=(\alpha\beta)$. The diagonalization of $C_{ab}(r,r')$ gives
\beq
 \sum_{b} \int C_{ab}(r,r')~\Psi^*_{A;b}(r') ~d\mu' ~=~ \lambda ~ \Psi^*_{A;a}(r)
\label{eigen}
\eeq
where the eigenvalues $\lambda$ are defined
\beq
\lambda = \sum_a  \int_D \Psi_{A;a}^{*}(r) \Psi_{A,a}(r) d\mu (r)
\label{41}
\eeq
In deriving this we used the fact the the wavefunctions (\ref{34}), (\ref{37}), (\ref{40}) are orthogonal to each other. 

We find that there are three distinct expressions for $\lambda$'s; one for each of the $SU(2)$ multiplets described above. After performing the angular integration in (\ref{41}) using (\ref{22})-(\ref{23}) we find
\beqar
\lambda_s^{(1)} & = & {(n+3)! \over {(s+3)! (n-1-s)!}} \int ^{R^2} { dx \over (1+x)^{n+4}} ~x^{s+3} \nonumber \\
\lambda_s^{(2)} & = & {(n+3)! \over {(s+2)! (n-1-s)! (n+1)}} \int ^{R^2} { dx \over (1+x)^{n+4}} \left[x^{s+2} + {s+1 \over {s+3}} ~x^{s+3} \right]
\label{42} \\
\lambda_s^{(3)} & = & {(n+3)! \over {(s+1)! (n-1-s)! (n+1) (n+2)}} \int ^{R^2} { dx \over (1+x)^{n+4}} \left[x^{s+1} + {2(s+1) \over {s+2}} ~x^{s+2} + {s+1 \over {s+3}}~ x^{s+3} \right] \nonumber
\eeqar
with the corresponding degeneracy $s+1$, $s+2$ and $s+3$. As $R^2 \rightarrow \infty$, $\lambda_s^{(I)} \rightarrow 1$ confirming the correct normalization for the wavefunctions. 

The expression for the entanglement entropy for the nonabelian lowest Landau level states for ${\mathbb {CP}}^2$ can now be written as 
\beqar
S &=&  \sum_{s=0}^{n-1} \left[ (s+1) H_s^{(1)} + (s+2) H_s^{(2)} + (s+3) H_s^{(3)} \right]  \nonumber \\
&& H_s^{(I)} =  -\lambda_s^{(I)} \log \lambda_s^{(I)} - (1 - \lambda_s^{(I)}) \log (1-\lambda_s^{(I)}) 
\label{43}
\eeqar
with the $\lambda^{(I)}$'s given in (\ref{42}). 

We will now show that $\lambda^{(I)}$'s in (\ref{42}) can be related to the abelian ones in (\ref{beta}) making a semiclassical calculation of (\ref{43}) similar to the abelian case. Making a change of variables to $t = x/(1+x)$ as before we find that (\ref{42}) can be written as 
\beqar
\lambda_s^{(1)} & = & {(n+3)! \over {(s+3)! (n-s-1)!}} \int ^{t_0}_0 dt ~ t^{s+3} (1-t)^{n-s-1}  \\
\lambda_s^{(2)} & = & {(n+3)! \over {(s+2)! (n-s-1)! (n+1)}} \int ^{t_0}_0 dt \left[ t^{s+2} (1-t)^{n-m-1}  -{{2 ~t^{s+3} (1-t)^{n-s-1}}\over {s+3}}   \right] \nonumber \\
\lambda_s^{(3)} & = & {(n+3)! \over {(s+1)! (n-s-1)! (n+1) (n+2)}} \int ^{t_0}_0  dt \Bigl[ t^{s+1} (1-t)^{n-s-1} - { {2 ~t^{s+2} (1-t)^{n-s-1}} \over {s+2}}  \nonumber \\
~~~~&&~~~~~~~~~~~~~~~~~~~~~~~~~~~~~~~~~~~~~~~~~~~~~~~~~~~~~~~~~~~~~~~~~~~+~{{2~t^{s+3} (1-t)^{n-s-1} }  \over {(s+2)(s+3)}}   \Bigr] \nonumber
\label{61a}
\eeqar
Comparing these to the abelian $\mathbb{CP}^k$ values (\ref{beta}) which we denote by $\lambda^{(\rm Ab)}$ we find the following relations,
\beqar
\lambda^{(1)}_{s, k=2} &=& \lambda^{(\rm Ab)}_{s+1, k=3} \nonumber \\
\lambda^{(2)}_{s, k=2} &=& { n+3 \over n+1} \lambda^{(\rm Ab)}_{s+1, k=2} -{2 \over n+1}  \lambda^{(\rm Ab)}_{s+1, k=3}  \\
\lambda^{(3)}_{s, k=2} &=& { n+3 \over n+1} \lambda^{(\rm Ab)}_{s+1, k=1} -{2 (n+3) \over {(n+1)(n+2)}}  \lambda^{(\rm Ab)}_{s+1, k=2}  +{2 \over{(n+1)(n+2)}} \lambda^{(\rm Ab)}_{s+1, k=3} \nonumber
\label{62b}
\eeqar
At the large $n$ limit the nonabelian eigenvalues $\lambda^{(I)}$ for $k=2$ coincide with the abelian ones for $k=1,2,3$ correspondingly,
\beqar
\lambda^{(1)}_{s, k=2} &=& \lambda^{(\rm Ab)}_{s+1, k=3} \nonumber \\
\lambda^{(2)}_{s, k=2} & \rightarrow &  \lambda^{(\rm Ab)}_{s+1, k=2}  \\
\lambda^{(3)}_{s, k=2} & \rightarrow &  \lambda^{(\rm Ab)}_{s+1, k=1}  \nonumber
\label{63b}
\eeqar
Similarly, at the large $n$ limit, the nonabelian entropy (\ref{43}) becomes a multiple of the abelian one in (\ref{60a})
\beqar
S &=&  \sum_{s=0}^{p} \left[ (s+1) H_{s,k=2}^{(1)} + (s+2) H_{s,k=2}^{(2)} + (s+3) H_{s,k=2}^{(3)} \right]  \nonumber \\
&\rightarrow & \sum_{s=0}^{p} \left[ (s+1) H_{s+1, k=3}^{(\rm Ab)} + (s+2) H_{s+1,k=2}^{(\rm Ab)} + (s+3) H_{s+1,k=1}^{(\rm Ab)} \right] \nonumber \\
& \rightarrow & 3~ n^{3/2} ~\pi ~( \ln 2)^{3/2}  ~{R^{3} \over {(1+R^2)^2}} = 3 ~S^{(\rm Ab)}
\label{64b}
\eeqar
The overall factor of 3 relating the nonabelian entanglement entropy to the abelian one above has to do with the fact that each lowest Landau state is an $SU(2)$ triplet, ${\rm dim} \tilde{J} = 3$. 
Although the calculation of the entropy in the case of a nonabelian background was explicitly done for $\mathbb{CP}^2$ and the triplet representation, one expects a more general statement to hold. In the large $n$ limit the degeneracy of the LLL in a case of a nonabelian background is \cite{KN,KN1}
\beq
N \sim {\rm dim} \tilde{J} ~{n^k \over k!}
\label{65}
\eeq
The corresponding phase-space volume in this case is $V_{\rm phase~space} = {\rm dim}\tilde{J}~{ n^k \over k!} \int d \mu$ and the corresponding phase-space surface area is 
\beq
A_{\rm phase~space} = n^{2k-1 \over 2}~  { {\rm dim}\tilde{J} \over k!} A_{\rm geom} = n^{k - \half} { {2 ~{\rm dim}\tilde{J}}   \over (k-1)!} ~{R^{2k-1} \over {(1+R^2)^k}} 
\label{66}
\eeq
Expressed in terms of the phase-space surface area the overall coefficient in the expression for the entanglement entropy is the same for {\it any abelian or nonabelian background} at large $n$
\beq
S \sim {\pi \over 2} ( \ln 2)^{3/2} ~A_{\rm phase~space}
\label{67}
\eeq

\section{Higher Landau levels} 

In this section we will consider the entropy for higher Landau levels focusing in particular on some of the differences in the behavior of the eigenvalues $\lambda$ between the lowest Landau level $q=0$, the first excited Landau level $q=1$ and the case of $\nu=2$ where both levels are filled. We will only consider the $k=1$ case, QHE on the sphere. Similar features apply for higher $k$. 

The wavefunctions for the $q$-th Landau level are of the form
\beq
\Psi^J_{m} (g) = \sqrt{N} \, \bra{J, m} g \ket{J, n} 
\label{68}
\eeq
where $J= {n/2} + q$ and ${\rm dim} J = n+2q+1$. The state $\ket{J, n} $ is not the lowest weight state of the $J$ representation. The lowest weight state is the LLL state with $n \rightarrow n+2q$. The $q=1$ states can therefore be generated by the action of $\hat{R}_{+}$ on the LLL states with $n \rightarrow n+2$. 
In the case of the sphere the representation of the $\hat{R}_i$ operators is of the form
\beqar
\hat{R}_{+} &= & -\epsilon_{\a\b} u^*_{\a} {\partial \over {\partial u_{\b}}} ~~,~~\hat{R}_{-} = \epsilon_{\a\b} u_{\a} {\partial \over {\partial u^*_{\b}}} \\
\hat{R}_{3}& =& \sum_{\alpha} {1 \over 2} [-u_{\a} {\partial \over {\partial u_{\a}}}+u^*_{\a} {\partial \over {\partial u^*_{\a}}} ]\nonumber
\label{69}
\eeqar
where
\beq
 u_{\alpha} =  {1 \over {\sqrt{1 + \bz z}}}  \left(\begin{matrix} z \\ 1 \\ \end{matrix} \right) 
\label{70}
\eeq
The $\hat{R}$-operators satisfy the $SU(2)$ algebra
\beq
[ \hat{R}_{+} ~,~\hat{R}_{-}] = 2 \hat{R}_3
\label{71}
\eeq
Based on the argument above the correctly normalized wavefunctions of the $q=1$ Landau level are
\beqar
\Psi_s^{q=1} &=& \sqrt{n+3} \sqrt{{(n+1)!} \over {s!(n+2-s)!}}~ \hat{R}_{+} ~\Bigl( u_1^s u_2^{n+2-s} \Bigr) \nonumber \\
&=& \sqrt{n+3} \sqrt{{(n+1)!} \over {s!(n+2-s)!}}  \Bigl[ {{-(n+2) \bar{z} z^s} \over {\sqrt{(1+z\bar{z})^{n+2}}}} + {s z^{s-1} \over \sqrt{(1+z\bar{z})^{n}} }\Bigr]
\label{72}
\eeqar
The corresponding eigenvalues of the two-point correlator are now of the form
\beq
\lambda_s^{(q=1)} =  {(n+3)! \over {(n+2)s! (n+2-s)!}} \int ^{R^2} dx { x^{s-1} \over (1+x)^{n+4}} ~[(n+2-s)x-s]^2
\label{73}
\eeq
Changing variables to $x={t \over (1-t)}$ as before we can rewrite the eigenvalues as
\beq
 \lambda_s^{(q=1)} = {{ (n+3)! (n+2)} \over {s! (n+2-s)!}}
 \int_0^{t_0} dt \, t^{s-1} (1-t)^{n-s+1} ~[t-{s \over {n+2}}]^2
 \label{74}
 \eeq
 The eigenvalue $\lambda_s^{(q=1)}$ as a function of $s$ is similar to $\lambda_s^{(q=0)}$ away from the transition region, but it displays a distinct step-like pattern around the transition $s =t_0~(n+2)$, as shown in Figures 5 and 6. The reason for this has to do with the fact that the wavefunctions (\ref{72}) have a node. Since they are generated by the action of $R_{+}$ on the LLL wavefunctions of monopole charge $n+2$ they are necessarily orthogonal to them. Since the LLL wavefunctions are nonzero and have no node, othogonality requires that the first level Landau wavefunctions must have a node.  Higher Landau level wavefunctions acquire more nodes and one expects more steps around the transition region for the corresponding eigenvalues $\lambda$. In fact based on the observation that the $q$-th level states can be written, up to normalization, as $\hat{R}^{q}_{+} | LLL, n \rightarrow n+2q > $ one can argue that the wavefunctions will have $q$ nodes and the profile of the corresponding $\lambda$ will display $q$ distinct steps. 
 A similar step-like pattern was observed in \cite{Dunne} for the higher Landau edge density functions for circular samples. 
 
\begin{figure}[h]
\begin{center}
\begin{minipage}{5cm}
\scalebox{.5}{\includegraphics{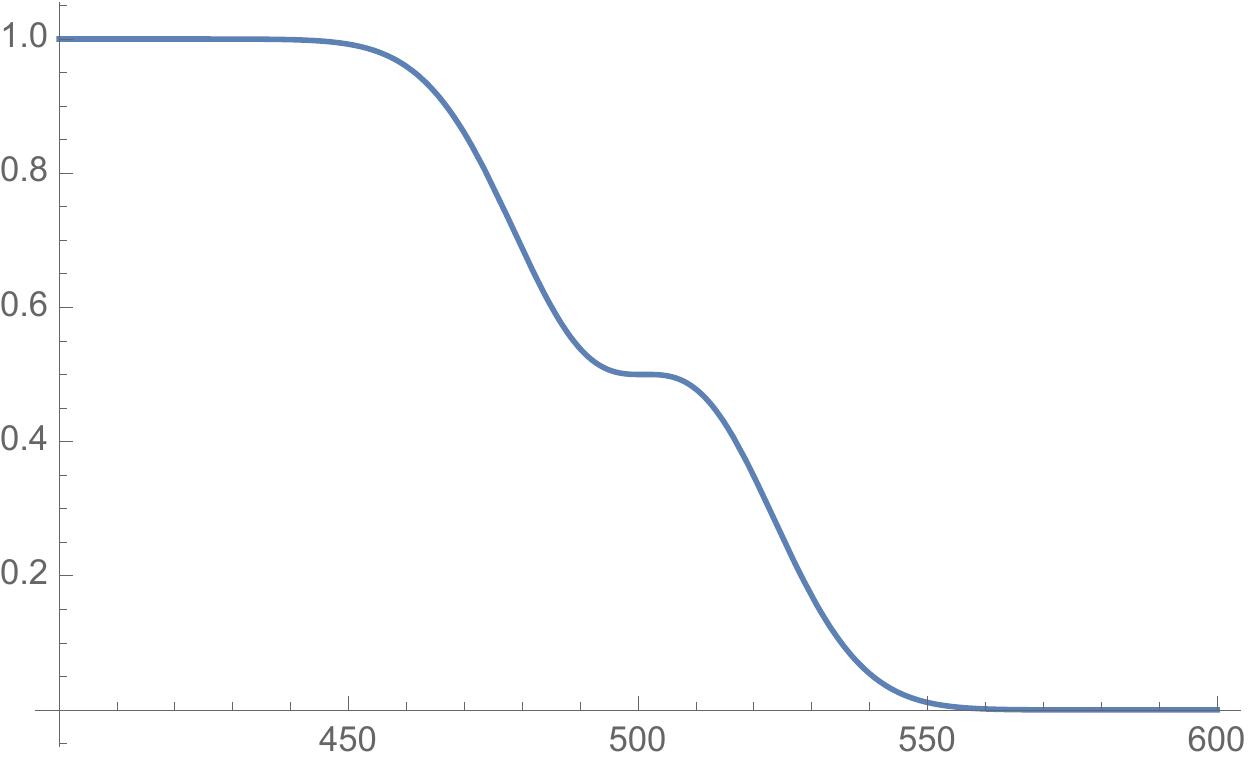}}
\caption{Plot of $\lambda_s^{(q=1)}$ as a function of $s$ for $k=1$, $n = 1000$ and $R=1$}
\label{g1}
\end{minipage}
\hskip 1in
\begin{minipage}{5cm}
\scalebox{.5}{\includegraphics{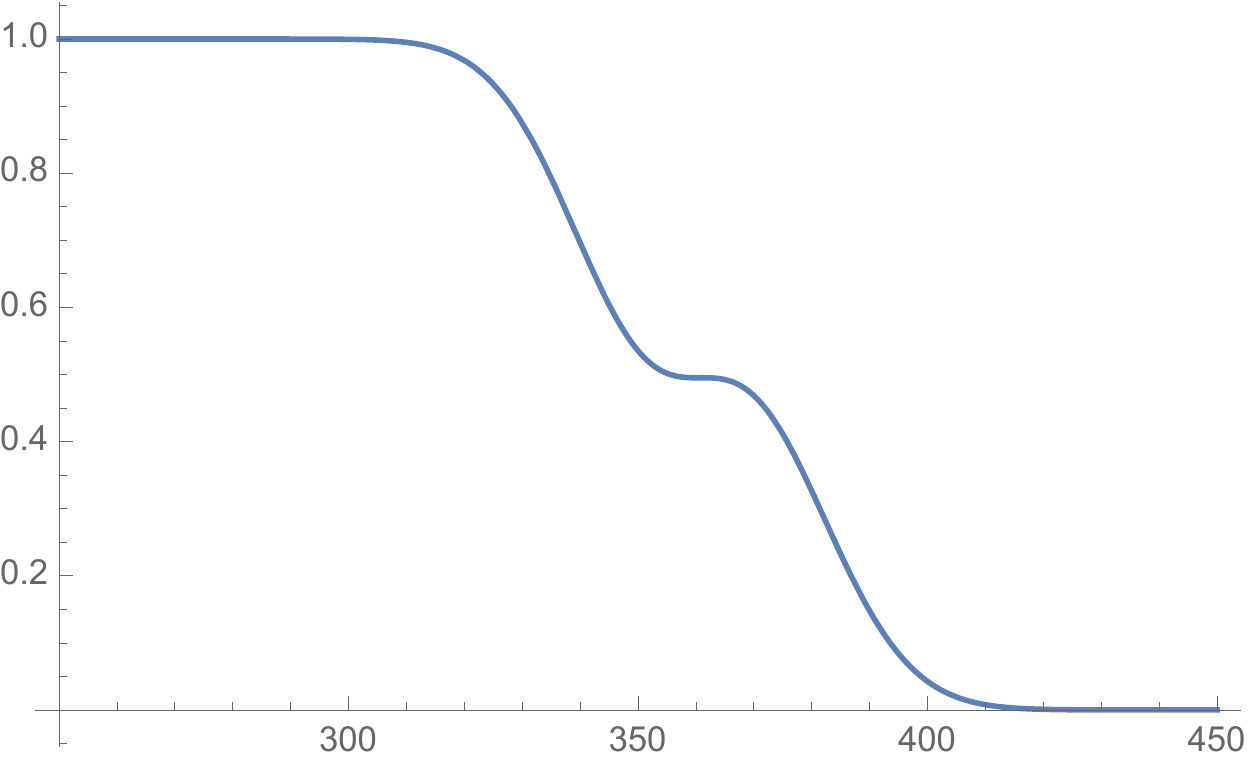} }
\caption{Plot of $\lambda_s^{(q=1)}$ as a function of $s$ for $k=1$, $n = 1000$ and $R=0.75$}
\label{g2}
\end{minipage}
\end{center}
\end{figure}
One can try to repeat the semiclassical analysis we did before for the first Landau level. 
Eq. (\ref{74}) can be written as 
\beqar
\lambda_s &=& {{ (n+3)! (n+2)} \over {s! (n+2-s)!}}
 \int_0^{t_0} dt \, e^{F(t)} ~[t-{s \over {n+2}}]^2   \label{90} \\
 F(t)&=&(s-1) \log t + (n-s+1) \log(1-t) \nonumber
\eeqar
The maximum of $F(t)$ occurs at $t^* = s-1/n$. Expanding $F(t)$ around $t^*$ we find that $e ^{F}$ becomes a Gaussian function of narrow width centered around $t^*$. In fact,
\beq
{d^2 F \over {dt^2}}\big\vert_{t^{*} }= - {n^3 \over {(n-s+1) (s-1)}} 
\label{91}
\eeq 
Around the transition region the main contribution of the integral comes from the range of $s$ around $s^*$ such that $t^*=t_0$, namely
\beq
t^*= {{s^*-1} \over {n}} = t_0 ~~~~\Rightarrow  s^* = t_0~n +1 ~~~~,~~~~n-s^*+1 = n(1-t_0)
\label{92}
\eeq
We now evaluate the integral in (\ref{90}) by expanding the integrand around $t_0$. In expanding $(t-s/(n+2))^2$ around $t_0$ we find that the large-$n$ contribution comes from the $(t-t_0)^2$ term. The constant and linear term in $t$ are suppressed by powers of $n$.
\beqar
\int_0^{t_0} e^{F(t)} (t-{s \over {n+2}})^2 & \sim &e^{F(t_0)} ~ \int_0^{t_0} \exp \left[ - {n\over 2 t_0 (1-t_0) }  (t-t_0)^2 \right] (t-t_0)^2 \nonumber \\
&=&e^{F(t_0)} {\sqrt{\pi} \over {4 n \sqrt{n}}} (2 t_0 (1-t_0))^{3/2}
\label{93}
\eeqar
Substituting this in (\ref{90}) and using Stirling's formula $n! = \sqrt{2 \pi n} ~(n/e)^n$, we find that 
\beq
\lambda_{s_*}^{(q=1)} = { 1 \over 2}
\label{94}
\eeq
independent of $t_0$ which is of course what is expected. 

Although the semiclassical treatment above is sufficient to capture the value of $\lambda^{(q=1)}$ at the transition point, the evaluation of $H_s^{(q=1)}$ is more involved since it cannot be approximated by a simple Gaussian due to the step like pattern for $\lambda^{(q=1)}$. 
$H_s^{(q=1)}$ will remain approximately flat in the step-like region, so higher derivatives around $s^*$ will be important to capture the correct behavior around the transition region. Figures 7 and 8 display the plots of $H_s^{(q=1)}$ around $\lambda=1/2$ based on the numerical evaluation of the exact expressions in (\ref{58}) and (\ref{74}). This clearly shows a deviation from the Gaussian distribution (see also Figure 13). 
\begin{figure}[t]
\begin{center}
\begin{minipage}{5cm}
\scalebox{.5}{\includegraphics{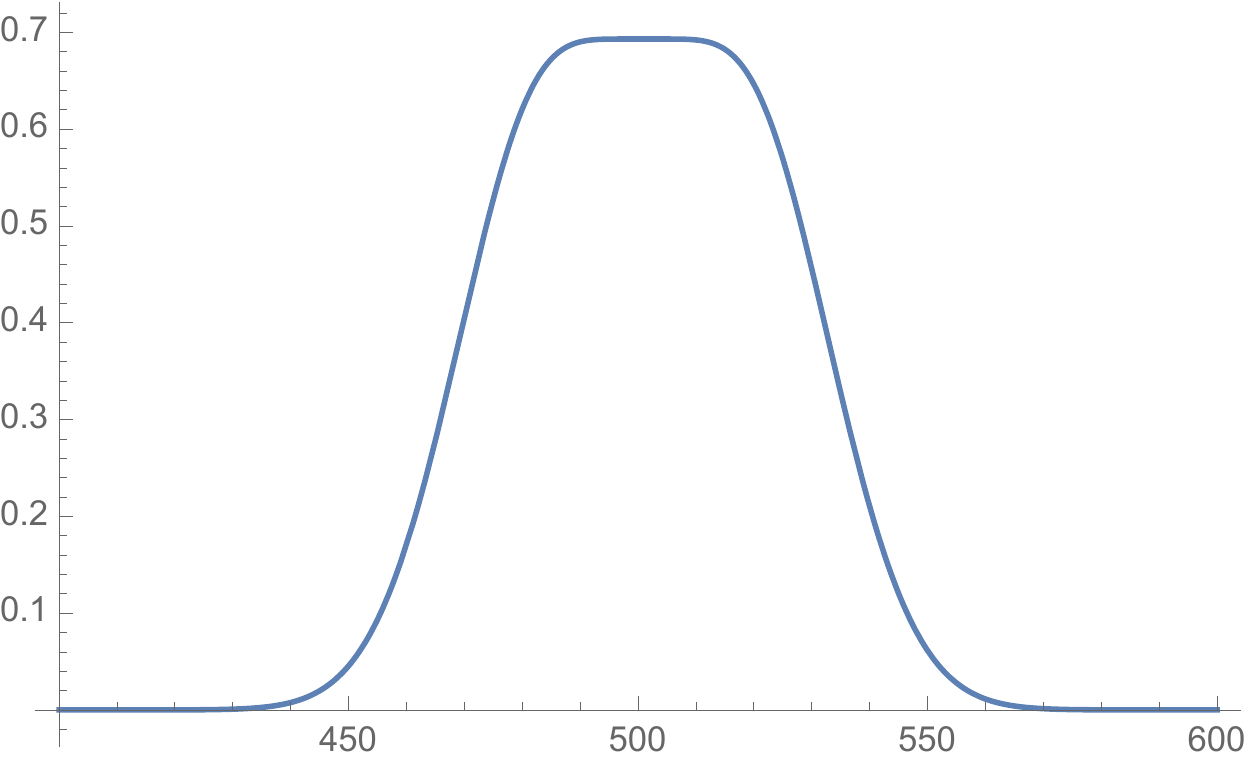}}
\caption{Plot of $H_s^{(q=1)}$ as a function of $s$ for  $n = 1000$ and $R=1$}
\label{g1}
\end{minipage}
\hskip 1in
\begin{minipage}{5cm}
\scalebox{.5}{\includegraphics{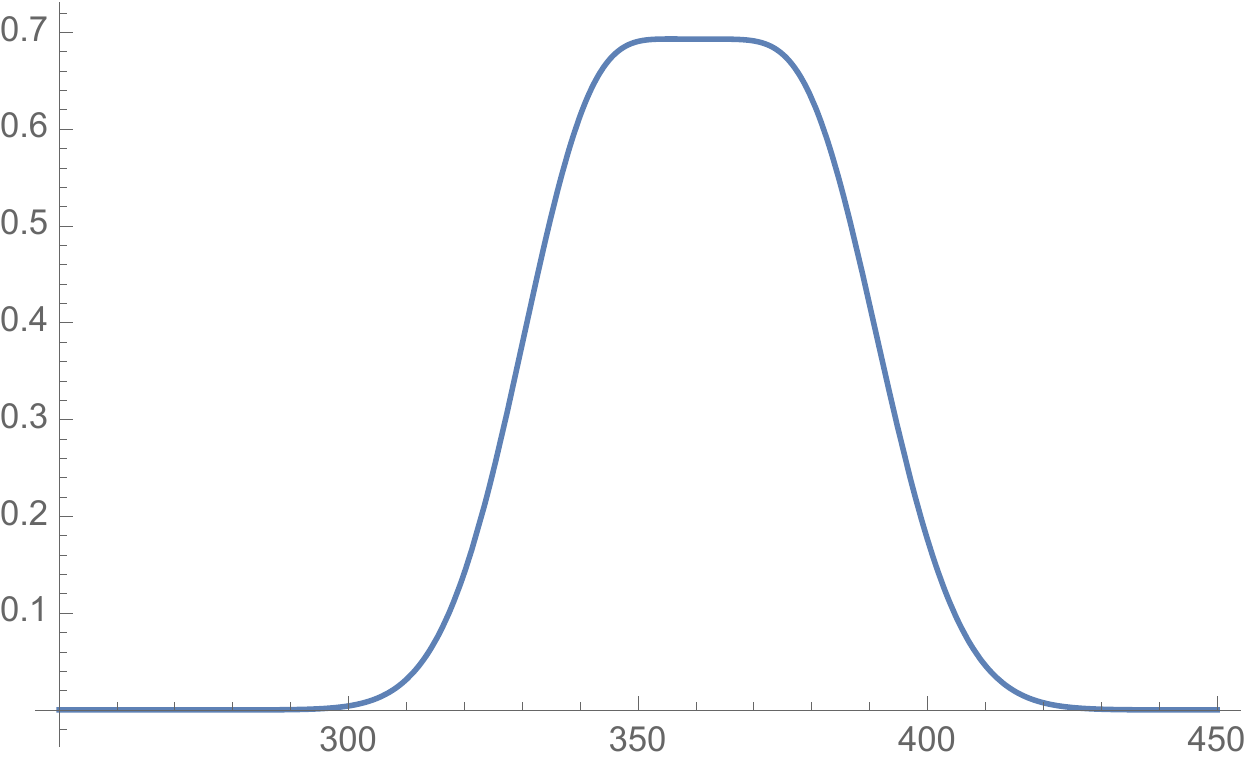} }
\caption{Plot of $H_s^{(q=1)}$ as a function of $s$ for  $n = 1000$ and $R=0.75$}
\label{g2}
\end{minipage}
\end{center}
\end{figure}
As a result the entropy for the first Landau level is larger than the entropy of the LLL even though the number of states are approximately the same at large $n$ ($n+1$ states for $q=0$ and $n+3$ states for $q=1$).
A numerical evaluation of the entropy shows that it obeys an area law and it gives
\beq
S^{(q=1)} =1.65 ~S^{(q=0)}
\label{82}
\eeq

When both $q=0$ and $q=1$ levels are filled, namely $\nu=2$, the situation is more involved as there are overlaps between the wavefunctions of different Landau levels. In particular,
\beqar
\delta\lambda_{s,s'} & = & \int_0^{R^2} \Psi^{*(q=0)}_s (r)~\Psi^{(q=1)}_{s'} (r) d\mu \nonumber \\
& = & \delta_{s+1,s'} ~{(n+1)! \over {s! (n-s)!}}\sqrt{{{n+3} \over {(s+1)(n+1-s)}}} \int_0^{R^2} \Bigl[ -{(n+2)x^{s+1} \over x^{n+3}}~+~{{(s+1)x^s} \over x^{n+2}} \Bigr]
\label{83}
\eeqar
The two-point correlator now is
\beq
C(r,r') = \sum_{s=0}^{n} \Psi^{*0}_s (r) \Psi^{0}_s (r') + \sum_{s=0}^{n+2} \Psi^{*1}_s (r) \Psi^{1}_s (r')
\label{84}
\eeq
and 
\beq
\int C(r,r') \left(\begin{matrix} \Psi_s^{*0}(r') \\ \Psi_{s+1}^{*1} (r') \\ \end{matrix} \right) ~d\mu'~=~\left( \begin{matrix} \lambda_s^{0}& \delta\lambda_{s,s+1} \\ \delta\lambda_{s,s+1} & \lambda_{s+1}^{1} \end{matrix} \right) \left(\begin{matrix} \Psi_s^{*0}(r) \\ \Psi_{s+1}^{*1} (r) \\ \end{matrix} \right)
\label{85}
\eeq
where $\lambda^{0}~,~\lambda^{1}$ are the eigenvalues we derived earlier for the lowest and first Landau level and $\delta\lambda$ is the overlap in (\ref{83}). 
There are $2n+4$ eigenvalues for the two-point correlator given by: $\lambda_0^1~, \tilde{\lambda}_{s}^{\pm} ~,~ \lambda^1_{n+2}$, where $s=0,\cdots,n$ and 
\beq
\tilde{\lambda}_s^ {\pm} = {{\lambda_s^0 + \lambda_{s+1}^1 \pm \sqrt{ (\lambda_s^0-\lambda_{s+1}^1)^2 + 4 (\delta\lambda)^2_{s,s+1} }}\over 2}
\label{86}
\eeq
The interesting feature here is that once both Landau levels are included the step like pattern in the profile of $\lambda^1$ disappears. The profile of the new $\tilde{\lambda}^{\pm}$ resembles that of $\lambda^0$ but shifted with respect to $\lambda^0$, see Figures 9 and 10. 
\begin{figure}[t]
\begin{center}
\begin{minipage}{5cm}
\scalebox{.5}{\includegraphics{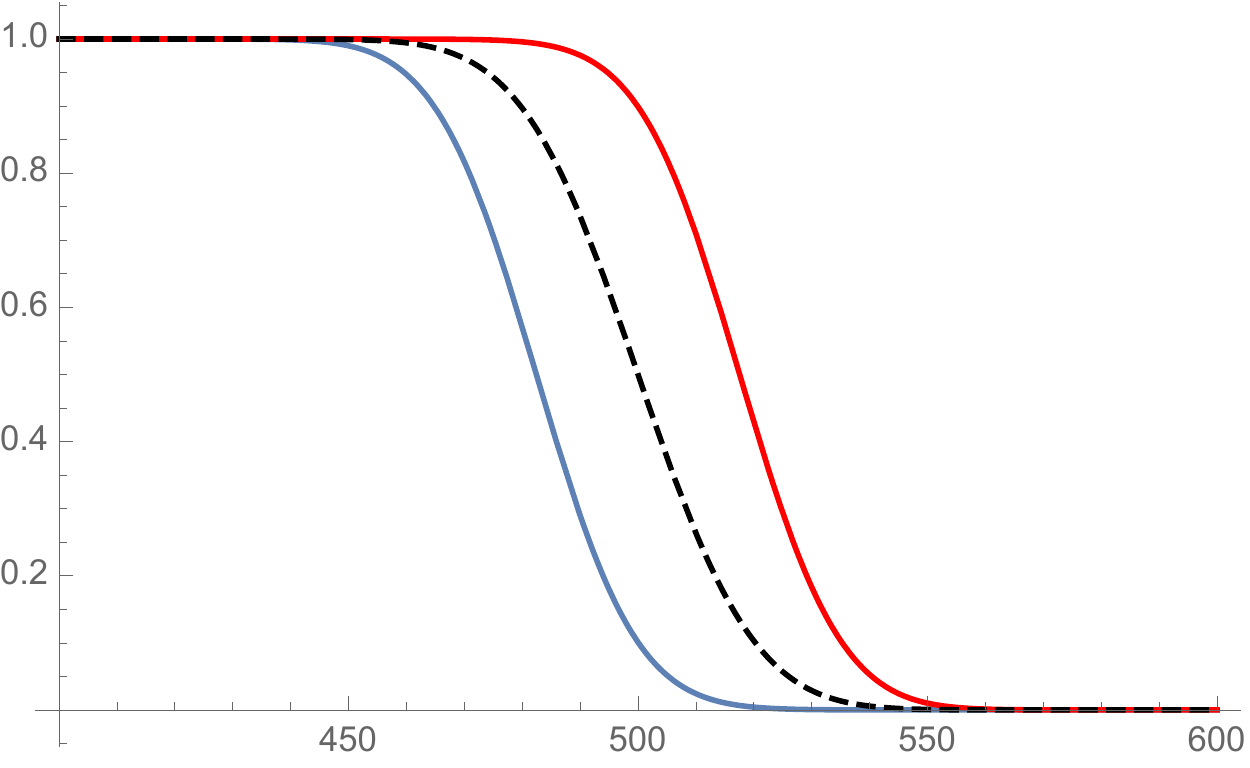}}
\caption{Plots of $\tilde{\lambda}^+$ (red to the right), $\tilde{\lambda}^-$ (blue to the left) as functions of $s$ for  $n = 1000$ and $R=1$, compared to $\lambda^0$ (dashed, center)}
\label{g1}
\end{minipage}
\hskip 1in
\begin{minipage}{5cm}
\scalebox{.5}{\includegraphics{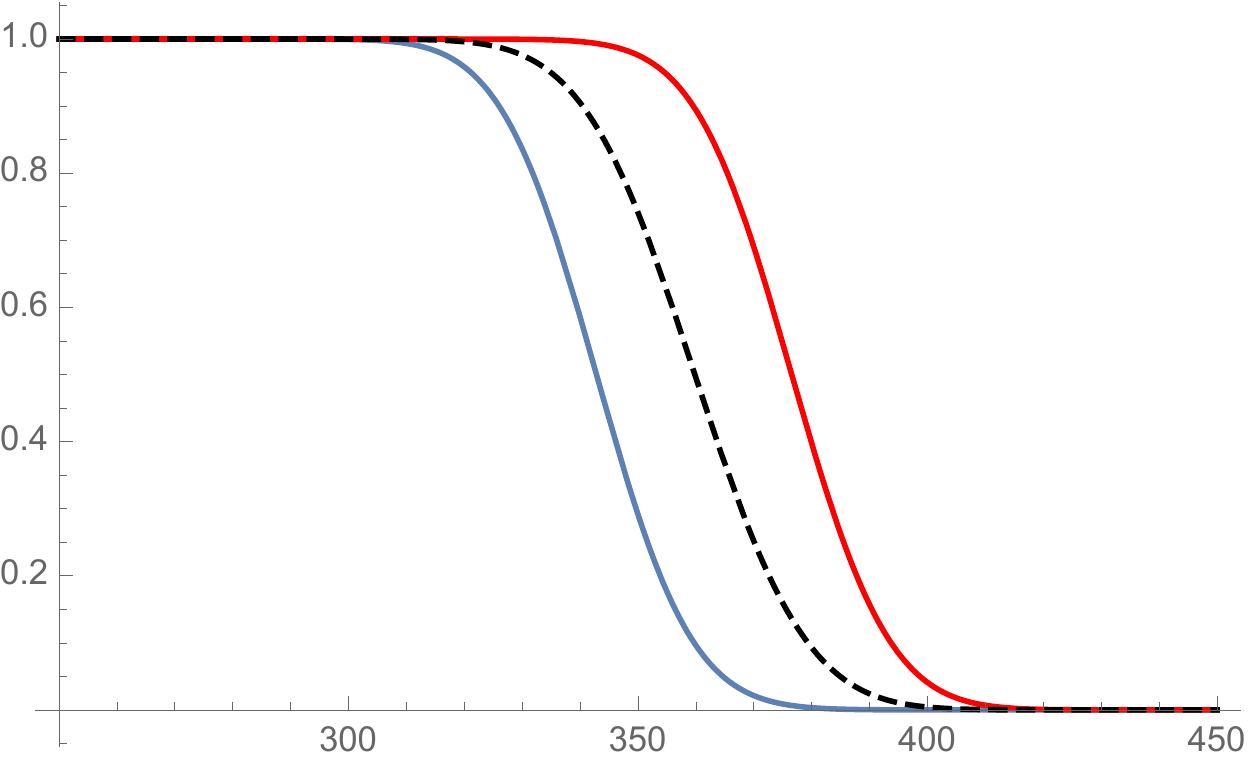} }
\caption{Plots of $\tilde{\lambda}^+$ (red to the right), $\tilde{\lambda}^-$ (blue to the left) as functions of $s$ for  $n = 1000$ and $R=0.75$, compared to $\lambda^0$ (dashed, center)}
\label{g2}
\end{minipage}
\end{center}
\end{figure}
As a result the corresponding entropy per mode $\tilde{H}^{\pm}_s$, where
\beq
\tilde{H}^{\pm}_s  = -\tilde{\lambda}^{\pm}_s \log \tilde{\lambda}^{\pm}_s - (1 - \tilde{\lambda}^{\pm}_s) \log (1-\tilde{\lambda}^{\pm}_s) \label{87}
\eeq
are Gaussian distributions each centered around the value of s for which $\tilde{\lambda}^{\pm} =1/2$ as shown in Figure 11 and 12.
\begin{figure}[h]
\begin{center}
\begin{minipage}{5cm}
\scalebox{.5}{\includegraphics{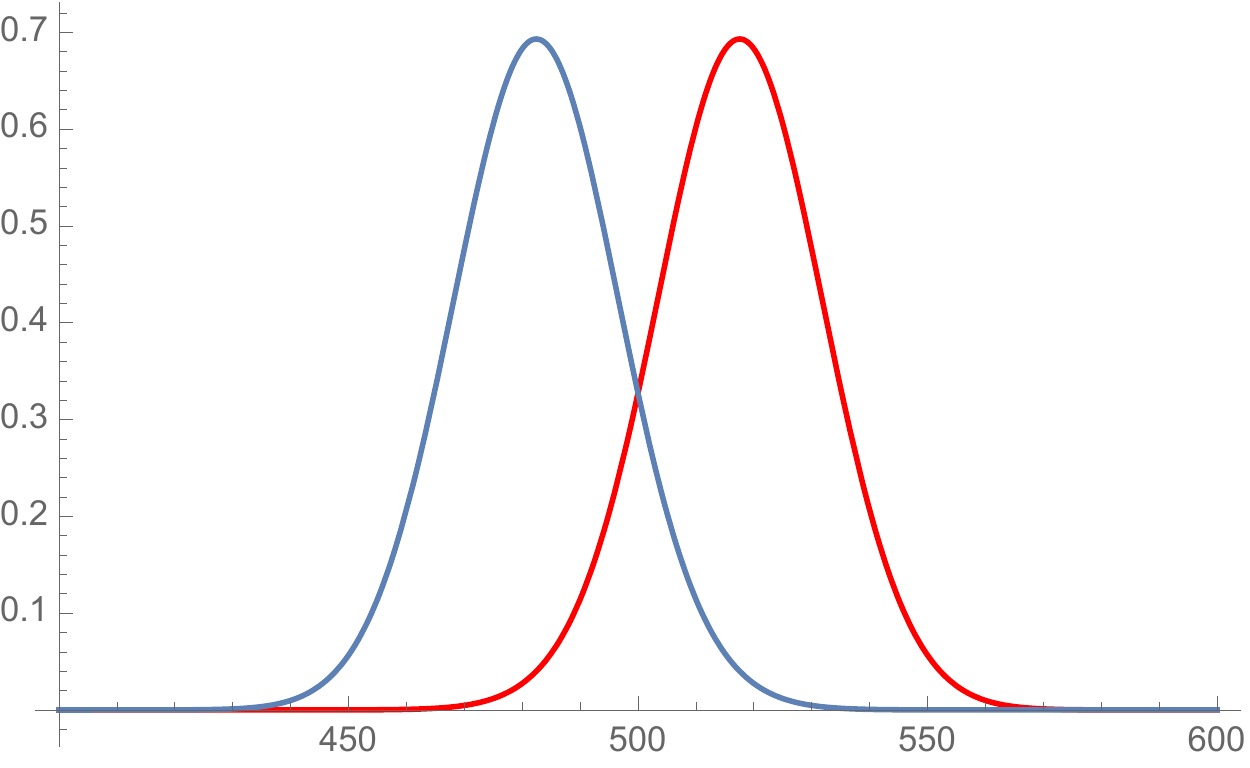}}
\caption{Plots of $\tilde{H}^+$ (red to the right) and $\tilde{H}^-$ (blue to the left) as functions of $s$ for  $n = 1000$ and $R=1$}
\label{g1}
\end{minipage}
\hskip 1in
\begin{minipage}{5cm}
\scalebox{.5}{\includegraphics{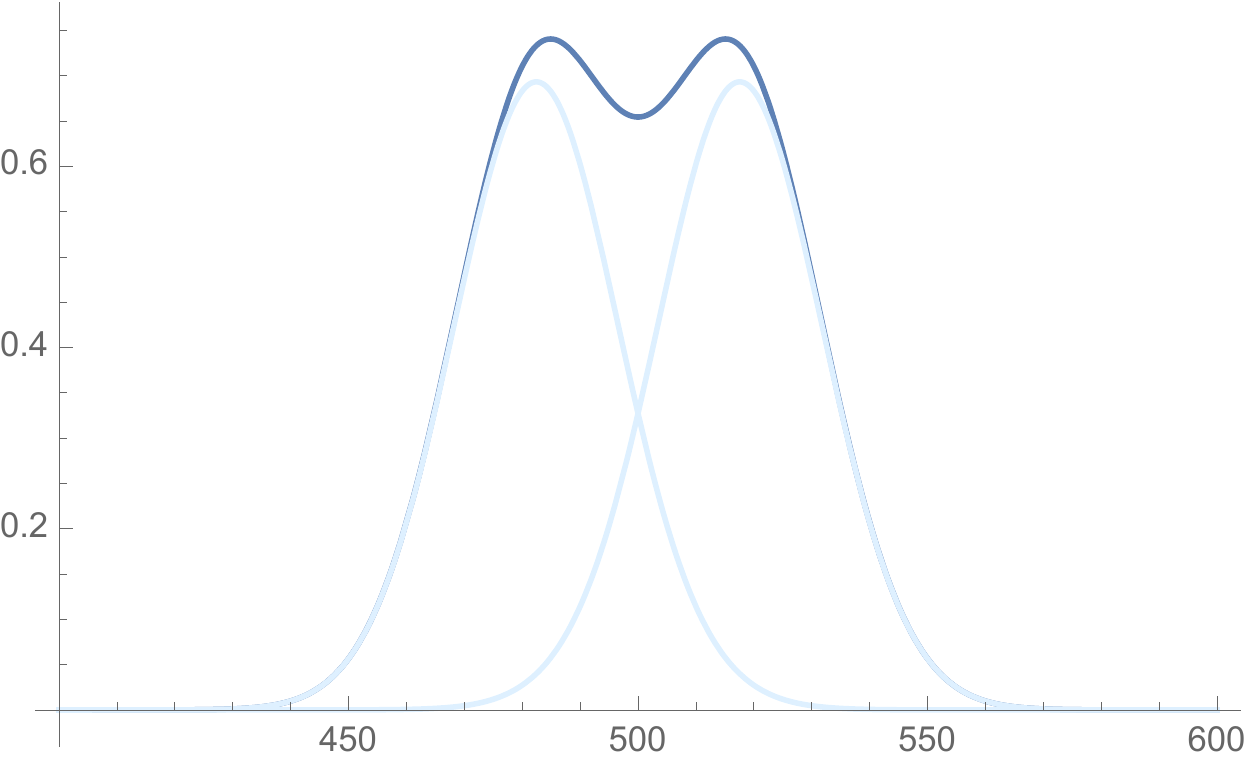} }
\caption{Plot of $\tilde{H}^+ + \tilde{H}^-$  as function of $s$ for  $n = 1000$ and $R=1$}
\label{g2}
\end{minipage}
\end{center}
\end{figure}
Figure 13 shows a comparison between $H^{(q=0)}~, H^{(q=1)}$ and $H^{(\nu=2)}$ which explains the differences in the values of the corresponding entropies, namely
\beq
S^{(\nu=2)} > S^{(q=1)} > S^{(\nu=1)}
\label{88}
\eeq
\begin{figure}[h]
\begin{center}
\scalebox{.5}{\includegraphics{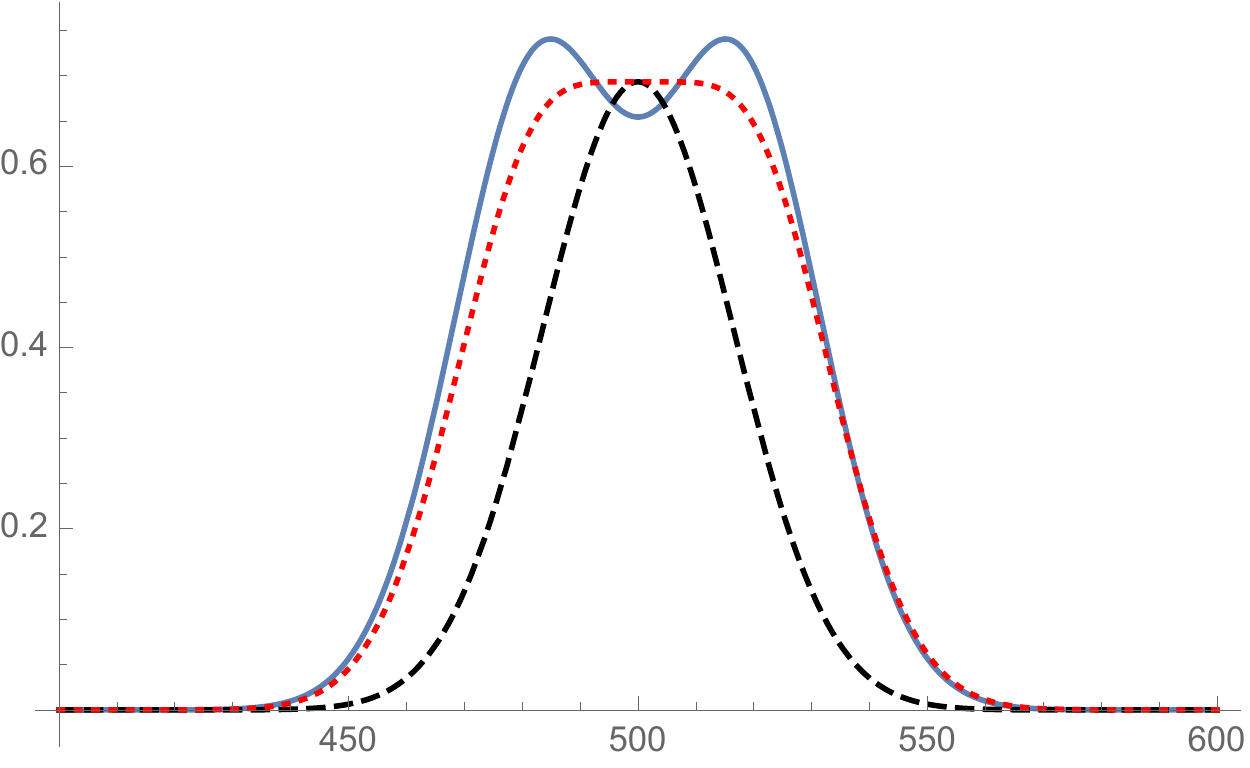}}
\caption{Plots of $H^{(\nu=1)}$ (black, dashed), $H^{(q=1)}$ (red, dotted) and $H^{(\nu=2)}$ (blue, solid) as functions of $s$ for  $n = 1000$ and $R=1$}
\label{g1}
\end{center}
\end{figure}

A numerical evaluation of the entropy for the $\nu=2$ case gives
\beq
S^{(\nu=2)} = 1.76~S^{(\nu=1)}
\label{89}
\eeq
This agrees with the result in \cite{Sierra}.

\section{Discussion} 

In this paper we have analyzed the entanglement entropy for fully filled $\nu=1$ higher dimensional quantum Hall effect on ${\mathbb {CP}}^k$ for abelian and nonabelian magnetic fields. The analytical calculation is based on a semiclassical analysis and we showed that the entropy satisfies the area law. In fact the entropy as expressed in terms of a phase-space entangling surface area has the same proportionality constant for all higher dimensions irrespective of the abelian or nonabelian nature of the background magnetic field. It will be interesting to see if a similar universal formula can be obtained for higher Landau levels. 

In the presence of edge degrees of freedom the entanglement entropy for the two-dimensional integer quantum Hall effect develops subleading logarithmic contributions \cite{other1}.  It has been shown in the two-dimensional $\nu=1$ quantum Hall effect that when the edge boundary intersects the boundary of the entangling surface there is an additional logarithmic contribution whose coefficient is determined by the central charge of the gapless edge modes \cite{estienne}- \cite{krempa}. In the context of higher dimensional quantum Hall effect we have previously analyzed the analogs of higher dimensional chiral abelian and nonabelian droplets, the edge spectrum and corresponding effective actions \cite{KN1}. It would be interesting therefore to extend the analysis of the entanglement entropy to these cases where the entangling surface and edge boundary overlap in higher dimensions and calculate the corresponding subleading corrections to the area law for the entanglement entropy. 

Similar considerations for higher Landau levels in both two and higher dimensions are also worth pursuing.

{\bf Acknowledgements}

I thank V.P. Nair for helpful discussions. This research was supported in part by the U.S.\ National Science
Foundation grant PHY-1915053 and by PSC-CUNY awards.


\end{document}